%% file: ms.tex
\documentclass[apj]{emulateapj}
\usepackage[colorlinks,linkcolor={blue},citecolor={blue},urlcolor={red}]{hyperref}
 \usepackage{epsfig} \usepackage{graphicx}
\usepackage{natbib} \usepackage{amsmath} \usepackage{amsfonts}
\usepackage{amssymb}
\usepackage{rotating}

\newcommand{\myemail}{linlin@shao.ac.cn}
\newcommand{\cliemail}{cli2015@tsinghua.edu.cn}

\newcommand{\nuvr}{NUV-$r$}
\newcommand{\concentration}{$R_{90}/R_{50}$}
\newcommand{\mstar}{M$_\ast$}
\newcommand{\lgmstar}{$\lg($\mstar$/$M$_\odot)$}

\newcommand{\dindex}{D$_n$(4000)} \newcommand{\ewhda}{EW(H$\delta_A$)}
\newcommand{\ewha}{EW(H$\alpha$)} \newcommand{\ha}{H$\alpha$}
\newcommand{\hd}{H$\delta$} \newcommand{\rt}{R$_{\mbox{t}}$}
\newcommand{\rbar}{R$_{\mbox{bar}}$}
\newcommand{\rdisk}{R$_{\mbox{disk}}$}
\newcommand{\rtrdisk}{\rt/\rdisk} \newcommand{\rtrbar}{\rt/\rbar}

\newcommand{\ddindex}{$\Delta$\dindex}
\newcommand{\hi}{H{\sc i}}
\newcommand{\mhi}{M$_{\rm HI}$}
\newcommand{\hifrac}{$\lg($\mhi$/$\mstar$)$}

\begin{document}

\title{Bar-induced central star formation as revealed by integral
  field spectroscopy from CALIFA}

\author{Lin Lin \altaffilmark{1},  Cheng Li \altaffilmark{2,1}, Yanqin
  He \altaffilmark{1,3}, Ting Xiao \altaffilmark{1}, Enci Wang
  \altaffilmark{1} }

\altaffiltext{1}{Shanghai Astronomical Observatory, Chinese Academy of
  Sciences, Shanghai 200030, China; \myemail} \altaffiltext{2}{Physics
  Department \& Tsinghua Center for Astrophysics, Tsinghua University,
  Beijing 100084, China; \cliemail} \altaffiltext{3}{Tianjin
  Astrophysics Center, Tianjin Normal University, Tianjin 300387,
  China}


\begin{abstract} 
We investigate the recent star formation history (SFH) in the inner
region of 57 nearly face-on spiral galaxies selected from the Calar
Alto Legacy Integral Field Area (CALIFA) survey. For each galaxy we
use the integral field spectroscopy from CALIFA to obtain
two-dimensional  maps  and radial profiles of three parameters that
are sensitive  indicators  of the recent SFH: the 4000\AA\  break
(\dindex), and the equivalent width of \hd\ absorption  (\ewhda) and
\ha\ emission (\ewha). We have also performed photometric
decomposition of bulge/bar/disk components based on SDSS optical
image. We identify a class of 17 ``turnover'' galaxies whose central
region present significant drop in \dindex, and most of them
correspondingly show a central upturn in \ewhda\ and \ewha.  This
indicates that the central region of the turnover galaxies has
experienced star formation in the past 1-2 Gyr, which makes the bulge
younger and more star-forming than surrounding regions. We find almost all
(15/17) the turnover galaxies are barred, while only half of the
barred galaxies in our sample (15/32) are classified as a turnover
galaxy. This finding provides strong evidence in support of the
theoretical expectation that the bar may drive gas from the disc
inward to trigger star formation in galaxy center, an important
channel for the growth/rejuvenation of pseudobulges in disc galaxies. 
\end{abstract}

\keywords{galaxies: spiral - galaxies: structure - galaxies: formation
  - galaxies: evolution - galaxies: stellar content }


\section{INTRODUCTION}
\label{sec:introduction}

Bars are commonly found in spiral galaxies. In the local universe, about 30\%
disc galaxies are barred, and the fraction  increases to 50-70\% if weak bars
are included or if NIR images are used for bar identification
\citep[e.g.][]{deVaucouleurs63, knapen00, marinova-jogee07, eskridge00,
masters11, lee-park12}. The bar fraction decreases  with increasing redshift,
dropping to about 20\% for all bars and below 10\% for strong bars at
$z\sim0.8$ \citep{sheth08}. In theory and $N$-body simulations bars play
crucial roles in driving the secular evolution of disc galaxies. Bars grow
through transfering angular momentum to the outer disk, or even to the
speriod/halo \citep{athanassoula03}. This process drives the gas in the disc
either outward to form ring-like structure in the outskirt, or inward to
trigger star formation in the central region \citep[e.g.][]{athanassoula92,
sellwood93, piner95, knapen00, sheth02, regan-teuben04, zurita-perez08}. The
gas inflow to the center is believed to make pseudobulges \citep[][and
reference therein]{kormendy04}.

Bar-induced gas inflow and the related central star formation have been
reported in many observational studies. Compared with unbarred galaxies, barred
galaxies are found to have higher gas concentrations \citep{sakamoto99,
jogee05, sheth05, regan06}, higher central star formation rates (SFRs)
\citep{dejong84, hawarden86, devereux87, puxley88, ho97}.  Flatter chemical
abundance gradients were found in barred galaxies from previous studies
\citep{zaritsky94, martin94}, while recent results shown no difference in
gas-phase \citep{sanchez14,sanchez-menguiano16} or stellar
\citep{cacho14,sanchez-blazquez14} metallicity gradients between galaxies with
or without bar.  A recent study by \citet{wang12} based on a volume-limited
sample of galaxies in the Sloan Digital Sky Survey \citep[SDSS;][]{york00}
found strong correlation between the central-to-global SFR and the presence of
a bar in face-on spiral galaxies: more than half of the galaxies with highly
concentrated SFR are barred. The authors suggested that the central star
formation of the other half galaxies which are unbarred may be triggered by
tidal interactions with companion galaxies, an idea which is supported by the
following two observational facts. On one hand, galaxy-galaxy interactions are
known to be able to trigger strong star formation in galaxy centers
\citep[e.g.][]{li08}. On the other hand, the presence of a bar in galaxies is
found to have weak/no correlation with galaxy-galaxy interactions \citep{li09,
lin14}.  These two findings combine to suggest that, the central star formation
as triggered by interactions and the central star formation induced by
bar-driven gas inflow are distinct events, and both are expected to contribute
to bulge growth.

Some studies suggested that bulges may be rejuvenated systems
\citep{thomas06, carollo07, obreja13, erwin15}. Old and young stellar
populations coexist in one bulge and these two populations are
kinematically distinguishable, with the old population having spheroid
kinematics and the secondary population being rotationally supported
\citep{emsellem01, perez09}. High frequency of young stellar
populations in bulges have been found in both barred galaxies
\citep{coelho11, mendez-abreu14} and close pairs \citep{kannappan04},
supporting both bars and interactions to be responsible for the
rejuvenation of the central bulge. 

Our understanding of the stellar populations and star formation
history  of different structural components in galaxies has improved
rapidly in recent years, thanks to the many integral field
spectroscopy (IFS) surveys.  These surveys have obtained spatially
resolved spectroscopy for samples of galaxies in the local Universe,
providing both two-dimensional maps and radial profiles of the stellar
population properties and kinematics across each galaxy.  For
instance, using data from the Calar Alto Legacy Integral Field Area
(CALIFA) survey \citep{sanchez12,sanchez16}, \citet{holmes15} detected
non-circular flows in 12 gas rich disc galaxies with intermediate
inclinations, and found 11 of them have a bar, providing strong
evidence in support of bars as the driver of gas flow and thus secular
galaxy evolution. Another nice example of IFS observations of barred
galaxies is presented in \citet{gadotti15}
which studied kinematics and stellar population content of NGC 4371,
a massive barred galaxy in the core of the Virgo Cluster, using data
from the Multi-Unit Spectroscopic Explorer (MUSE), an IFS instrument
recently commissioned at the VLT. The MUSE data revealed an inner disc
and a nuclear ring, which are rotationally supported and dominated by
stars older than 10 Gyr. This suggested the bar in the galaxy formed
at $z\sim1.8$, and thus may have an extended  impact on the galaxy
evolution over a long time.

In this work we make use of data from the second data release of the
CALIFA survey \citep{garciabenito15} and study the recent star
formation history of the inner region for 57 nearly face-on spiral
galaxies. In particular, we aim to improve our understanding of the
physical link between the central star formation history and the
presence of the bar structure. For this purpose we measure three
spectral indices: \dindex\ (the break at around 4000\AA\ in optical
spectra of galaxies), \ewhda\ (equivalent width of the \hd\ absorption
line) and \ewha\ (equivalent width of the \ha\ emission line), which
are known to be sensitive indicators of stellar populations formed at
different times in the past 1-2 Gyr. Therefore one may have an
indication of the recent star formation history by combining the three
parameters \citep[e.g.][]{bruzual03, kauffmann03, li15}. As we will
show, quite a large fraction of our galaxies show clear signature of
recent/ongoing star formation in the central region and almost all
these galaxies are barred. On the other hand, interestingly, the bar
structure is presented in only half of the galaxies that show the
central star formation. 

We will first describe the CALIFA sample and our methodology of
measuring the SFH diagnostic parameters (\S\ref{sec:data}), and then
present our results in \S\ref{sec:results}.  We discuss the
implications of our results on bar-driven gas inflow and bulge
growth/rejuvenation in \S\ref{sec:discussion}, and summarize our
conclusions in \S\ref{sec:conclusions}.

\section{Data} 
\label{sec:data}

\subsection{The CALIFA survey}
\label{sec:califa_survey}

We use data from the second data release (DR2; \citet{garciabenito15})
of the Calar Alto Legacy Integral Field Area (CALIFA) survey
\citep{sanchez12,sanchez16}, which is obtaining integral field
spectroscopy (IFS) for a sample of 600 nearby galaxies at
$0.005<z<0.03$. The target galaxies are selected according to their
angular size so that each galaxy is covered to more than two times its
effective radii ($R_e$) by the CALIFA integral field unit (IFU) which
has a fixed field of view of 65$\arcsec$$\times$72$\arcsec$.  The
spatial resolution of the IFU is about 390 pc per arcsec at the median
redshift (z$\sim$0.02) of the sample, thus providing well resolved
spectroscopy suitable for studying the bulge and bar structure of
galaxies in the current work.  For each galaxy CALIFA obtained  IFS
with two different gratings: one with a lower spectral resolution of
R$\sim500$ and  a wider wavelength coverage of 3745$-$7500
\AA\ (V500), and one with  a higher resolution of R$\sim1200$ and a
narrower wavelength range  of 3650$-$4600 \AA\ (V1200).  In this work
we opt for the V500 data cubes only, in order to have measurements of
the H$\alpha$ emission line. 

\subsection{Sample selection}
\label{sec:sample_selection}

The CALIFA DR2 includes 200 galaxies, a random subset of the full
target sample. The data was released in September, 2014 when we
started this work. For each galaxy, in addition to the spectroscopic
data cube, a Hubble morphological type accompanies the data release,
determined by the CALIFA team based on visual examination of the
optical image from the Sloan Digital Sky Survey (SDSS;
\citet{york00}).  Details of the classification can be found in
\citet{walcher14}.

\begin{figure}[t!] 
\centering  \includegraphics[width=0.5\textwidth]{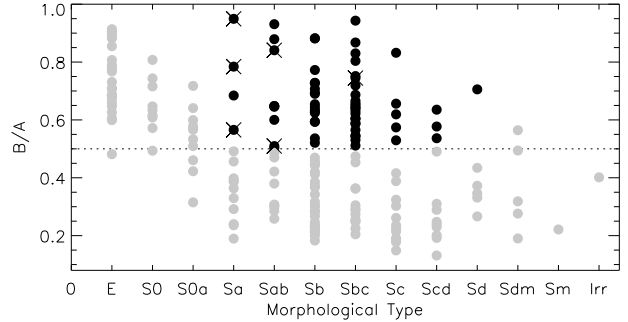}
\caption{Distribution of morphological type and minor-to-major axis
  ratio ($b/a$) for the CALIFA DR2 sample. The dashed line represents
  $b/a$=0.5, the divider we adopt to select face-on galaxies.  The 57
  face-on spiral galaxies studied in this paper are highlighted as
  solid black dots. The crosses indicate the galaxies with  obvious
  dust-lane or merging systems, and these are excluded from our
  analysis.  }
\label{fig:sample_select}
\end{figure}

Figure~\ref{fig:sample_select} shows the CALIFA/DR2 galaxies in the plane of
minor-to-major axis ratio ($b/a$) versus Hubble type, where $a$ and $b$ are the
semi-major and semi-minor axis lengths determined from SDSS $r$-band image by
the growth curve analysis \citep{walcher14}.  As can be seen, the sample covers
all morphological types, and a wide range of inclination as well. In order to
perform reliable photometric decomposition (see below), we restrict ourselves
to galaxies with $b/a$ $>$ 0.5, thus an inclination angle less than
60$^{\circ}$.  Furthermore, we've excluded galaxies with elliptical, lenticular
or irregular morphology, as well as three interacting/merging systems ({\sl
NGC~5394}, {\sl NGC~5614}, {\sl NGC~5630}), limiting the analysis to normal
spiral galaxies with regular, undisturbed morphology\footnote{We keep one
galaxy ({\sl NGC~3106}) which display slightly disturbed image in the optical,
but still show quite regular morphology. See Section 4.1 for further
discussion.}.  Finally, we exclude {\sl NGC~1056}, {\sl NGC~6314},  and {\sl
NGC~7615} because of their obvious dust lane across the optical image.

These restrictions give rise to a sample of 57 face-on spiral
galaxies, of which 31 have a strong bar in the center, 9 to have a
weak bar and 17 to have no obvious bar-like structure, according to
the visual classification by \citet{walcher14}. These galaxies are
highlighted as solid/dark circles in Figure~\ref{fig:sample_select}.

\subsection{Full spectral fitting} 
\label{sec:spectral_fitting}

For each spectrum in the V500 datacubes, we have performed a
decomposition of the emission-line component and the stellar component
(the continuum plus absorption lines) using the public code {\tt
  STARLIGHT} \citep{cidfernandes04}. With the emission lines being
masked out, the stellar component in the observed  spectrum is fitted
with a model spectrum which adds up a base of 45 single stellar
populations (SSPs) evenly distributed on an age-metallicity grid, with
15 ages ranging from 1 Myr to 13 Gyr and 3 different metallicities
($Z=0.01, 0.02, 0.05$).  The SSPs are selected from the models of
\citet{bruzual03},  which were built based on the stellar library
STELIB  compiled by Le Borgne et al. (2003), the Padova 1994
evolutionary tracks,  and a Chabrier (2003) initial mass function
(IMF).  The {\tt STARLIGHT} compares the model spectra with the
observed one and searches for the best-fit model through the
$\chi^2$-minimum scheme and the Markov Chains Monte Carlo (MCMC)
algorithm.  Intrinsic stellar extinction is modeled as rising from the
foreground dust, with the extinction law of \citet{cardelli89}. More
details about {\tt STARLIGHT} and the BC03 models can be found in the
relevant papers mentioned above.  The observed spectra are corrected
for Galactic extinction and resampled to have an wavelength interval
of $\Delta\lambda=2$\AA, before the spectral fitting. The spectral
fitting is performed to each spaxel separately, without Voronoi
binning. The typical S/N per spaxel depends on radius. The typical S/N 
of the stellar continuum in the central region, Re and 2Re are around 40, 
20 and 10, respectively. The observed spectra are well fitted, with a reduced
$\chi^2$ of 1-2 in general.

Following \citet{li15} and previous studies
\citep[e.g.][]{kauffmann03},  we use D$_{n}$(4000) and the equivalent
width of H$\delta$ absorption  line, EW(H$\delta_{A}$), as indicators
of the recent star formation history. It is well known that
D$_{n}$(4000) is a good proxy for the stellar age below $\sim2$ Gyr,
and H$\delta_{A}$ a sensitive tracer of the star formation that
happened 0.1-1 Gyr ago \citep{bruzual03}. In addition, we use the
equivalent width of H$\alpha$ emission line, EW(H$\alpha$), as an
indicator of the current star formation activity.  For each spaxel in
the data cubes we measure D$_{n}$(4000) and EW(H$\delta_{A}$) from the
best-fit spectrum, adopting the definition of the two parameters
introduced in \citet{balogh99} and \citet{worthey97}. EW(H$\alpha$) is
measured by fitting a single Gaussian profile to the pure emission
line  spectrum, obtained by subtracting the best-fit stellar component
from the observed one. 

We'd like to point out that, although {\tt STARLIGHT} provides
quantitative measurements of mean stellar age, in the rest of this
paper we will mostly use the three diagnostic parameters, that is,
D$_{n}$(4000), EW(H$\delta_A$) and EW(H$\alpha$). In particular, we
will use relative measurements of the three indices and concentrate on
their radial profiles and gradients.  In most cases we do not present
the derived measurements of stellar age, to avoid potential effect of
the known issues in current stellar population synthesis models, such
as the degeneracy between age, metallicity and dust, and the
uncertainty in IMF. However, as we will show at the end of the paper,
our conclusions based on the three indices are in good agreement with
those from the stellar age measurements. The typical errors for 
D$_{n}$(4000), EW(H$\delta_{A}$) and lg(EW(H$\alpha$)) in spaxels of
SNR$=$10 are 0.11, 1.05 \AA and 0.15 dex, respectively.

\begin{figure}[t!]
\centering \includegraphics[width=0.5\textwidth]{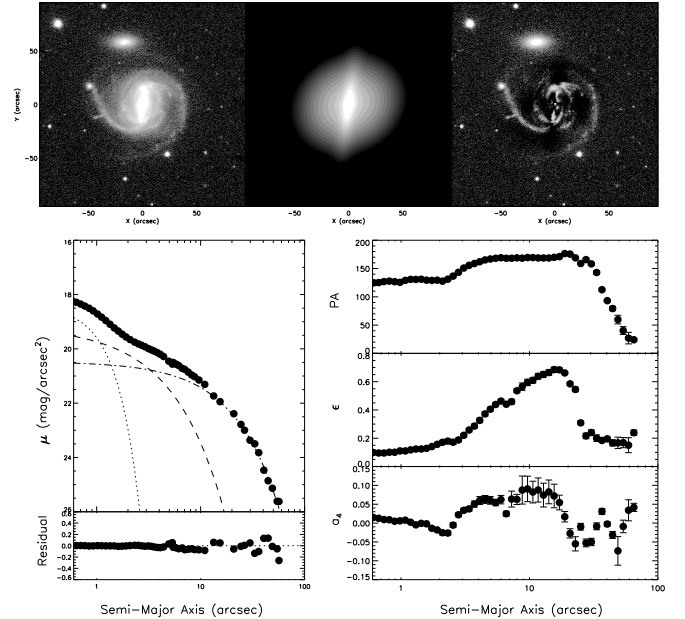}
\caption{An example of photometric decomposition ({\sl NGC~5000}). The
  top three panels (from left to right) show the original SDSS r-band
  image, the model image and the residual from GALFIT. The bottom left
  panel shows the one-dimensional surface brightness profiles (upper)
  and the data-to-model ration (lower). The dotted, dashed and
  dash-dotted lines represent the bulge/bar/disk components of the
  GALFIT model. The bottom-right panels show the position angle, the
  ellipticity and the $a_4$ parameter given by the IRAF/ELLIPSE
  analysis.  See text for more details. }
\label{fig:galfit_example}
\end{figure}

\begin{figure*}[t!]
\centering  \epsfig{figure=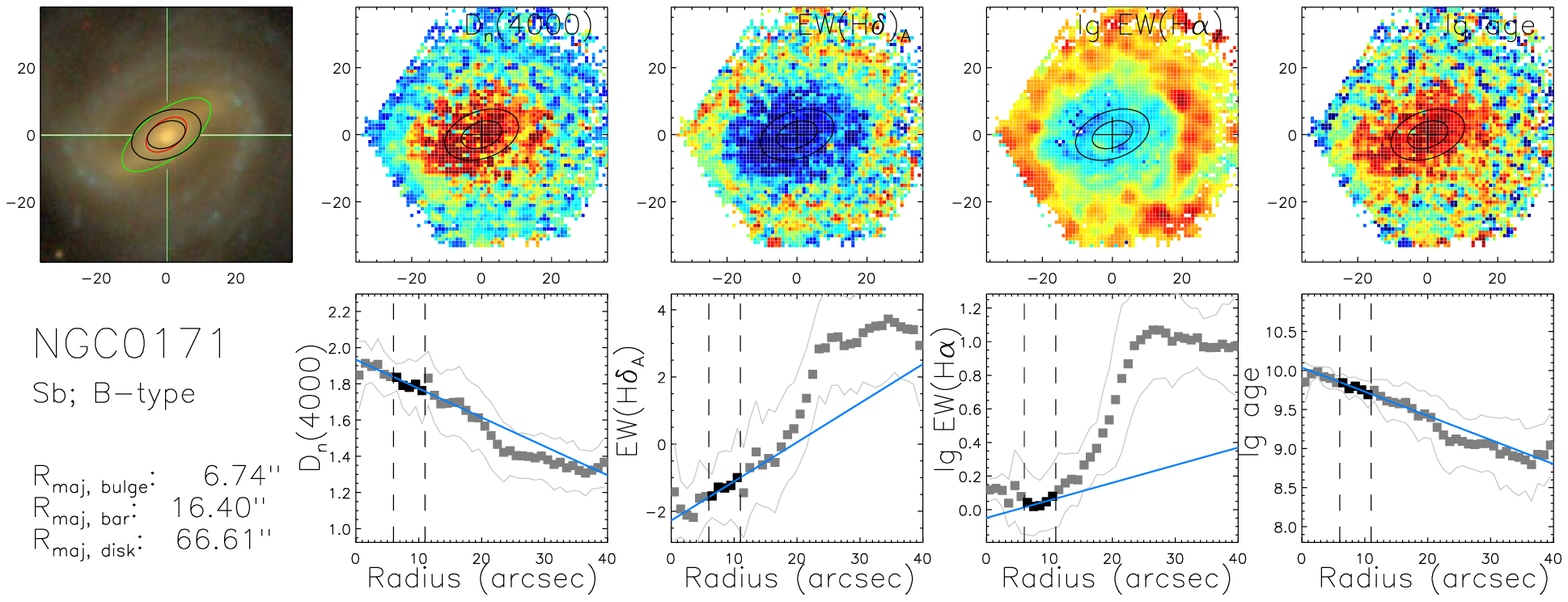,width=0.8\textwidth}
\epsfig{figure=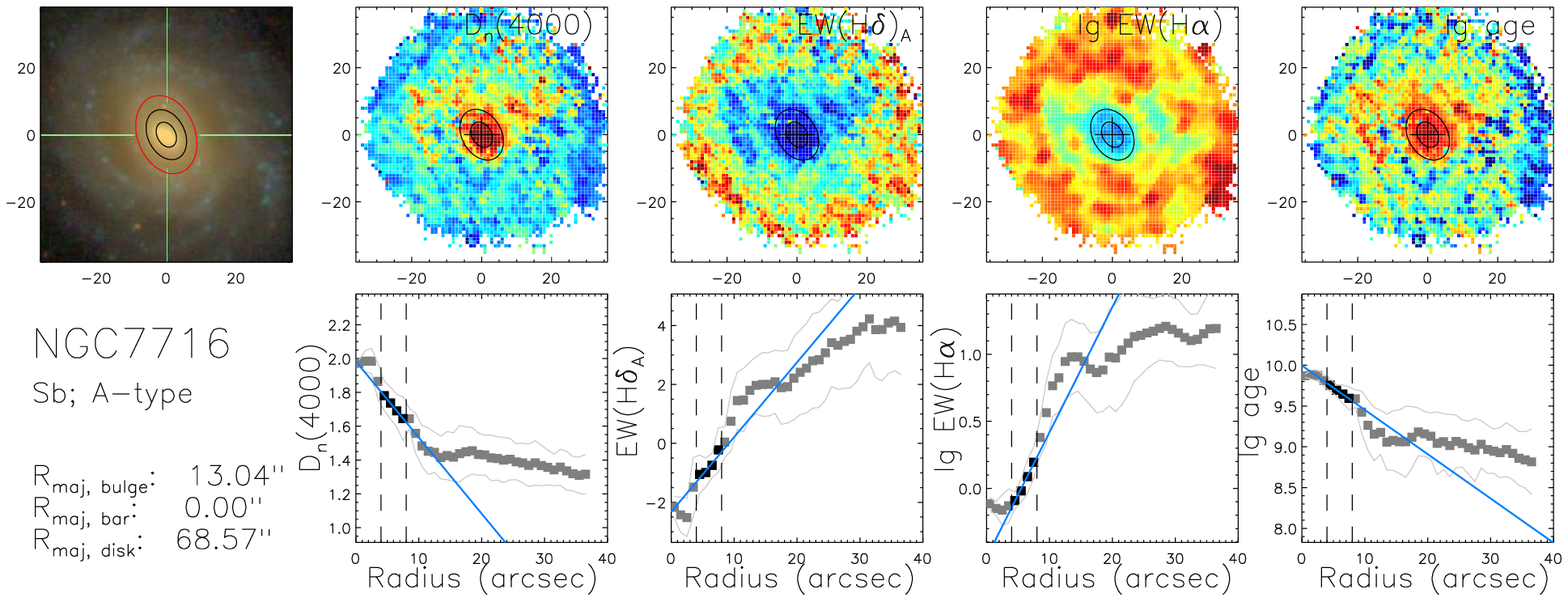,width=0.8\textwidth}
\epsfig{figure=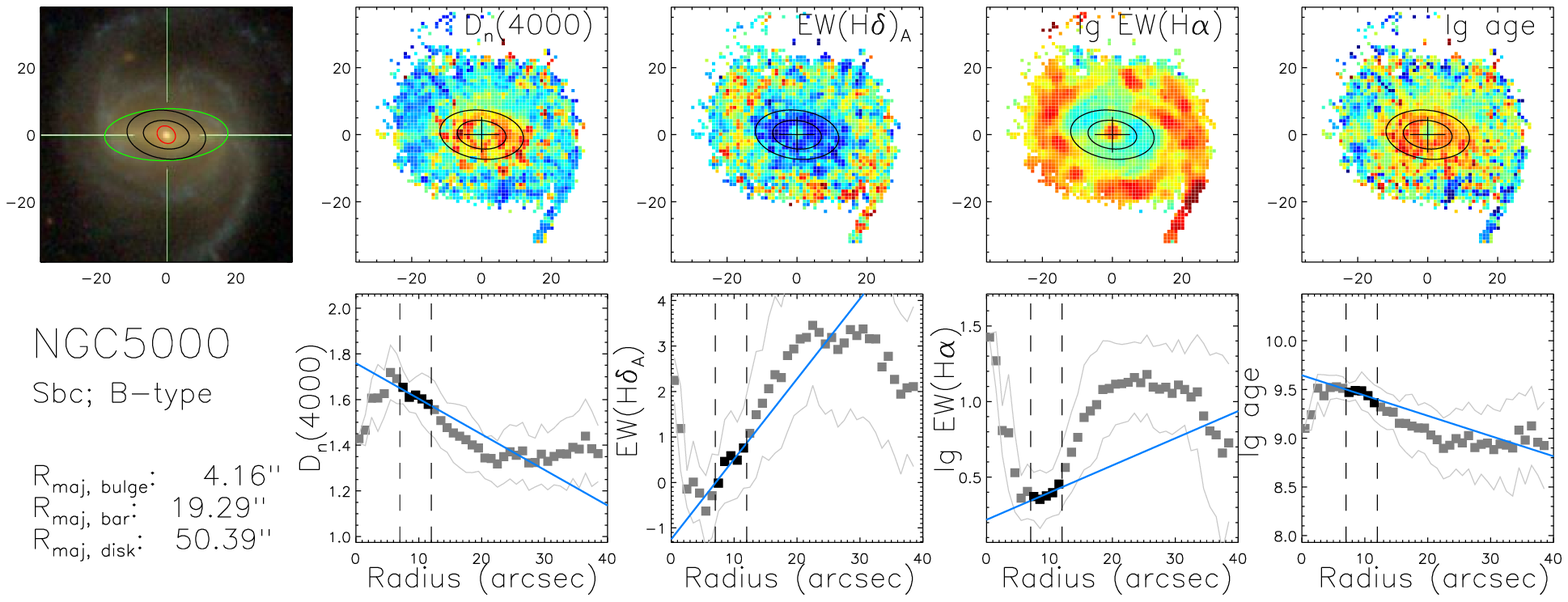,width=0.8\textwidth}
\epsfig{figure=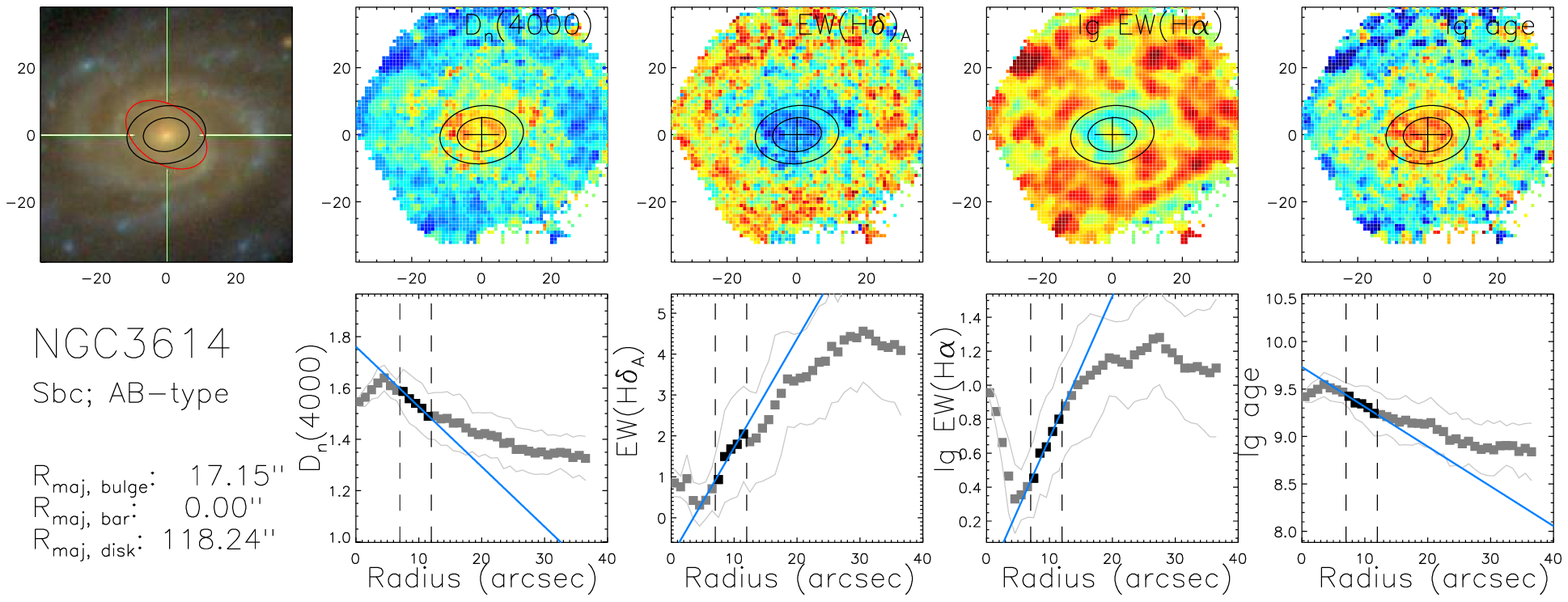,width=0.8\textwidth}
\caption{Optical $gri$ image (left column) and maps/profiles of
  \dindex, \ewhda, \ewha\ (column 2-4) and stellar age (right column)
  are shown for four example galaxies in our sample (from top to
  bottom): {\sl NGC 0171}, {\sl NGC 7716}, {\sl NGC 5000} and {\sl NGC
    3614}. In each image the red and green ellipse indicates the
  radius of the bulge and (if available) bar. The two black ellipses
  indicate the radial range in which we perform the linear fit to the
  radial profile, as also indicated as the two vertical dashed lines
  in the panels of the radial profiles, where the linear fit is
  plotted as the blue line.  }
\label{fig:map_examples}
\end{figure*}

\subsection{Photometric decomposition }  
\label{sec:photometry}

We have performed a careful, photometric decomposition of the bulge
and disk components for all the 57 face-on spirals in our sample.  For
barred galaxies we also include the bar structure as a third component
in the decomposition. The decomposition is performed on the SDSS
$r$-band image. We download the original images from the SDSS DR7 Data
Archive Server,  and we determine the sky background and mask out
foreground stars and neighboring sources, in the same way as described
in \citet{he13}.

As mentioned above, visual identification of bars is done for the
CALIFA galaxies. For this work we re-do the identification by applying
the IRAF task {\tt ELLIPSE} to the background subtracted image of each
galaxy in our sample. The {\tt ELLIPSE} fits  elliptical annuli to
surface brightness isophotes at  logarithmically increased radii, and
simultaneously determines a one-dimensional radial profile for the
ellipticity, the  position angle of the major axis, and the surface
brightness. Following \citet{barazza08}, we identify the bar structure
according to the presence of an abrupt drop in the ellipticity
profile:  a galaxy is classified to have a bar if the ellipticity
increases above 0.25 as one goes from galactic center outwards  and
then drops by at least 0.1 at some radius. As a result, 32 of the  57
galaxies in our sample are classified as barred, of which 29  are
visually identified as barred (B-type) and 3 as unknown type (AB-type)
in the CALIFA/DR2. We will use our classification for the rest of this
paper. Indeed we have repeated all the analysis using the
classification from the CALIFA/DR2 and found all our conclusions to
remain the same. 

For each galaxy we perform a two-dimensional decomposition by applying
the public code {\tt GALFIT} \citep{peng02} to the $r$-band SDSS
image, with either a two component model  (bulge+disc) if the galaxy
doesn't have a bar, or a three component model (bulge+bar+disc) if it
is barred.  The disc component is characterized by an exponential
profile, while  both the bulge and bar components are described by  a
S\'ersic profile. The model is convolved with the point spread
function determined from the observed image, in order to take into
account the effect of seeing. Parameters of the different components
of the best-fit model are listed in Table \ref{tbl1}.  

As an example, Figure \ref{fig:galfit_example} shows the observed, best-fit and
data-to-model residual images for one of our galaxies, {\sl NGC~5000},
as well as the one-dimensional profiles of the bulge, disc and  bar
components from the model. The {\tt ELLIPSE}-based profiles  are also
shown, which are used to identify the presence of the  bar component
in the galaxy. In the last column of Table~\ref{tbl1} we list the
reduced $\chi^2$ of the model fits, which are mostly at $\sim1-2$
indicating that the images are well fitted with these models.

From the best-fit models, we determine the sizes of bulge/bar/disk as
the radius at which the surface brightness is equal to
25.0 mag arcsec$^{-2}$. These radii will be used for the analysis
below.
 
\section{RESULTS} 
\label{sec:results}

\subsection{2D maps and radial profiles of \dindex, \ewhda\ and \ewha} 
\label{sec:maps} 

For each of the 57 galaxies in our sample, we first obtain the
two-dimensional (2D) maps of the  three diagnostic parameters,
\dindex, \ewhda\ and \ewha, from which we then estimate  the radial
profiles of the same parameters. When constructing these profiles, we
have corrected for the effects of inclination on the radius for each
spaxel in the datacubes, using the global ellipticity, the position
angle of the major axis and the major-to-minor axis ratio  from the
CALIFA data release \citep{walcher14}, and adopting a constant spatial
sampling of 1\arcsec\ along the major axis and fixed ellipticity and
position angle for all the radial bins. In
Figure~\ref{fig:map_examples} we present the 2D maps and the radial
profiles for four example galaxies, of which two are barred galaxies
({\sl NGC~0171}, {\sl NGC~5000}) and two are unbarred ({\sl NGC~7716},
{\sl NGC~3614}). We also show the SDSS $gri$ image and the
maps/profiles of stellar age given by {\tt STARLIGHT}.

An overall impression from the 2D maps is that all the galaxies in our
sample display spatial variation in all the three parameters, though
to varying degrees and to differing radial extents. We find that, in
general, the galaxies in our sample show larger \dindex, as well as
smaller \ewhda\ and \ewha\ in the inner region than in the outer
regions. These general trends are consistent with what have been found
for spiral galaxies from optical imaging, where a spiral galaxy
typically displays a red, dense bulge in the center, surrounded by
bluer, more diffuse light in/between spirals at larger radii. These
results also echo the previous studies of the stellar population
gradients based on stellar population synthesis applied to the CALIFA
datacubes \citep[e.g.][]{perez09, sanchez-blazquez11}.

When compared to the 2D maps, the radial profiles reveal these trends
in a more compact and quantitative manner: \dindex\ typically shows a
negative slope over the entire galaxy, while both \ewhda\ and
\ewha\ show positive slopes. We see that the profile of \dindex\ can
be divided into two parts in most cases: an inner part which can be
described by a linearly decreasing function, and an outer part which
is also linear but usually flatter than the inner part. The
\ewhda\ and \ewha\ profiles can be similarly divided into two parts
with a linear inner part, but the outer part displays more
fluctuations. The dividing radius between the two parts changes from
galaxy to galaxy, but it appears to occur at the radius which
separates the central bulge/bar from the outer spirals. The fluctuant
outer profiles of \ewhda\ and \ewha\ are consistent with both the
clumpy, highly structured  outskirt of their 2D maps, and the
spiral/clumpy light distribution in the optical image. In contrast,
the map of \dindex\ looks pretty smooth in the same regions.  As
indicators of current and very recent star formation, both \ewha\ and
\ewhda\ are expected to be strong only in strong star forming regions
(H{\sc ii} regions) which are discretely distributed in the spirals of
galaxies. The  \dindex\ indicates relatively older stellar
populations, formed over the past 1-2 Gyr, so it is not surprising to
see a smoother distribution of this parameter.

\begin{figure*}[t!]
\centering  \includegraphics[width=0.33\textwidth]{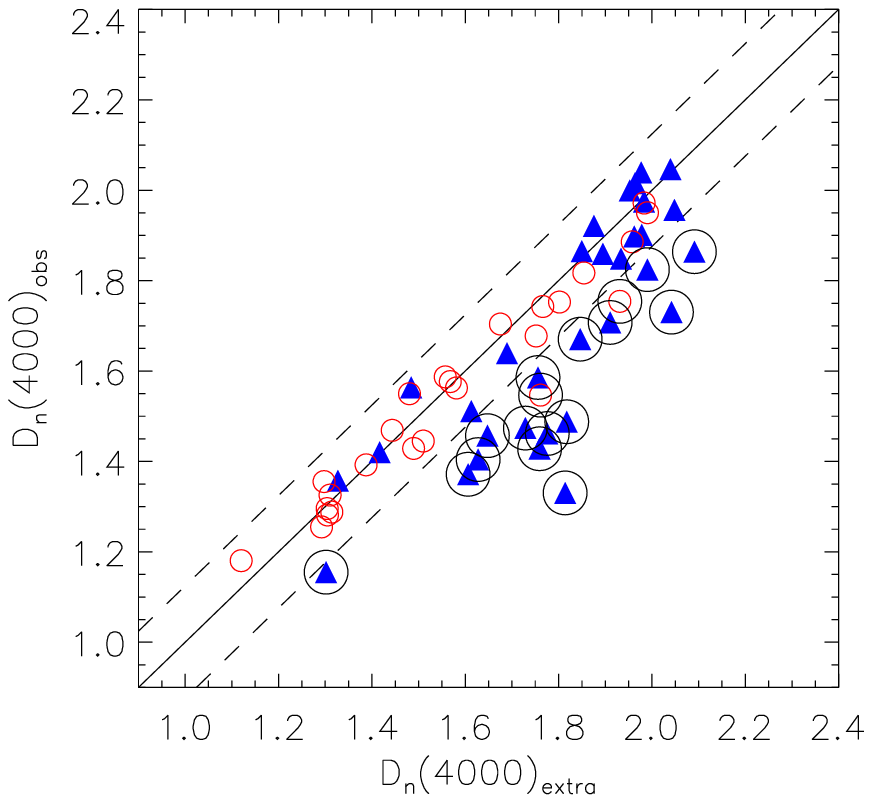}
\includegraphics[width=0.33\textwidth]{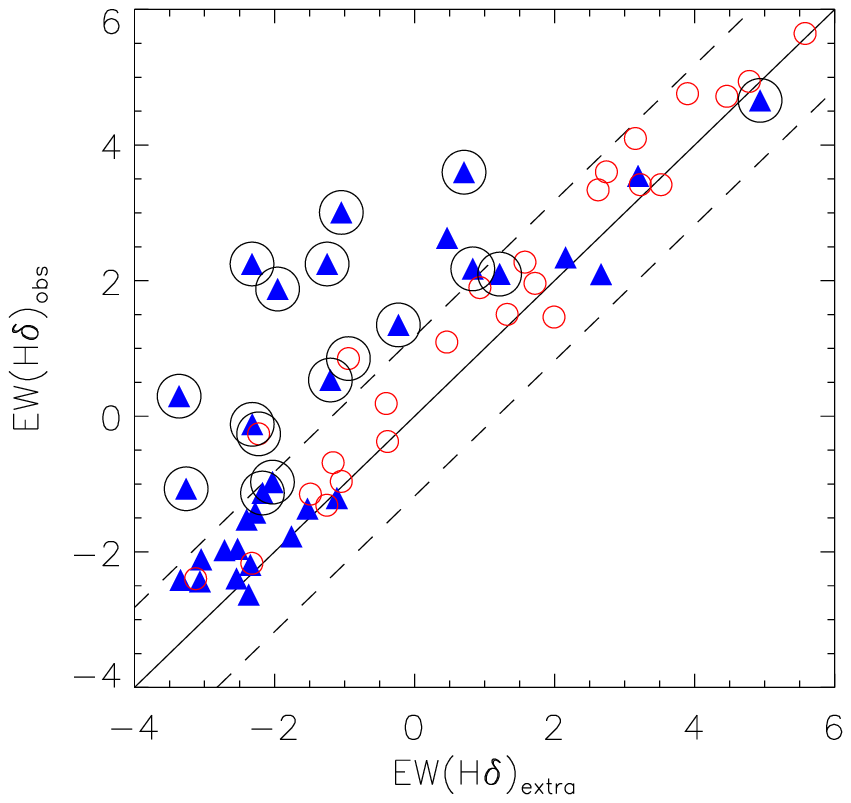}
\includegraphics[width=0.33\textwidth]{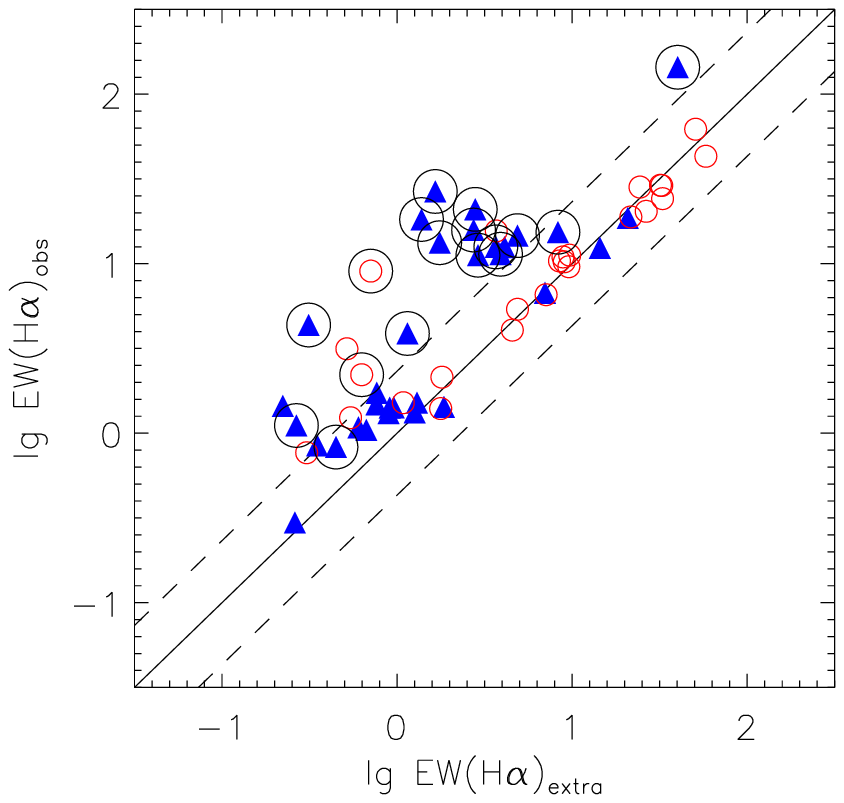}
\caption{Comparison between the observed central values and the inward
  extrapolated values for the three diagnostic parameters (from left
  to right): \dindex, \ewhda\ and \ewha. Barred and unbarred galaxies
		are plotted separately as solid blue triangles and red open circles.
  The turnover galaxies identified by the \dindex\ profile are
  highlighted with a big black circle. The solid line is the $1:1$
  relation, and the two dashed lines indicate the 1$\sigma$ scatter of
  all the galaxies around the $1:1$ relation.  }
\label{fig:obs_vs_extra}
\end{figure*}

\subsection{The central ``turnover'' in \dindex, \ewhda\ and \ewha\ profiles}
\label{sec:turnover}

In this subsection we will concentrate on the inner region of our
galaxies where a bulge and (sometimes) a bar are identified, aiming to
examine the recent star formation history in the central bulge, and in
particular the role of the bar structure in this regard.

As described above, in most of our galaxies the inner region can be
described by a linear profile, with a negative radial gradient in
\dindex\ and positive gradients in \ewhda\ and \ewha. In some of our
galaxies, thanks to its high spatial resolution, the CALIFA data
reveals a ``turnover'' in the inner-most region, where the three
diagnostic parameters significantly deviate from the featureless,
linear profile. This can be seen from the maps/profiles of {\sl
  NGC~5000} and {\sl NGC~3614}, the lower two galaxies shown in
Figure~\ref{fig:map_examples}.  In both galaxies, the profile of
\dindex\ has a negative slope at radii larger than $\sim$5\arcsec, but
exhibits a significant drop at smaller radii. Similarly, an upturn is
seen in both \ewhda\ and \ewha, also in the inner-most region of the
two galaxies. The other two galaxies, {\sl NGC~0171} and {\sl NGC~7716}
as shown in the upper two rows of the same figure, do not show a
significant ``turnover'' in either \dindex\ or \ewhda, whereas a
similar but weaker upturn is noticeable in \ewha, also in the very
center of the galaxy.

In order to quantify the ``turnover'' feature, for each galaxy, we
first determine an outer radius of its central region, $R_{out}$,
according to the photometric decomposition described in
\ref{sec:photometry}. We define $R_{out}$ as the radius along major
axis  where the $r$-band surface brightness $\mu_r$ falls below 25 mag
arcsec$^{-2}$, in either the best-fit bulge component (for unbarred
galaxies) or the best-fit bar component (for barred galaxies)
\footnote{We note that, at $R_{out}$, the bar component (if
  applicable) is exclusively much stronger than the bulge component,
  and using the total surface brightness of both components instead of
  the bar component alone doesn't change $R_{out}$, nor the analyses
  in the rest of the paper.}.  We also determine an inner radius,
$R_{in}$, which is set to $3\arcsec$ for unbarred galaxies, or
$R_{25,bulge}$ if the galaxy has a bar. Here, $R_{25,bulge}$ is the
radius along major axis at which the $r$-band surface brightness of
the bulge component reaches 25 mag arcsec$^{-2}$. We then fit a linear
function to the observed profiles of all the three diagnostic
parameters over the radial range $(R_{in}+R_{out})/2\pm2\arcsec$,
i.e. using the middle 5 data points in the radial range of
$R_{in}<R<R_{out}$. Actually we have also done the fitting using all
the data points falling in between $R_{in}$ and $R_{out}$, but finding
the two radii to cover a part of the spiral arms or the central
turnover region in some cases. In practice, we find that the middle 5
data points provide a good enough fit to the overall profile in the
central region, with little contaminations from the transition
regions. For comparison we have performed the linear fit also to the
stellar age profile derived from {\tt STARLIGHT}. In
Figure~\ref{fig:map_examples} the linear fits are shown as a blue
solid line for each profile. The 5 data points used for the fitting
are highlighted as black squares, and the two vertical dashed lines
indicate the lower and upper radii of these data points.

The ``turnover'' of a given parameter is then quantified by comparing
the observed value in the central spaxel with the value at $R=0$ as
extrapolated from the best-fit linear
profile. Figure~\ref{fig:obs_vs_extra} presents this comparison for
the three parameters. In each panel, the black solid line indicates
the $1:1$ relation, and the two dashed lines the $1-\sigma$ scatter
(0.14) of the galaxies around this relation. We plot barred and
unbarred galaxies as blue triangles and red circles separately. The
differing behavior of the two classes is striking. The majority of the
unbarred galaxies follow well the $1:1$ relation, whereas a large
fraction of the barred galaxies deviate significantly from that
relation, i.e., falling below the lower dashed line. Specifically, the
fraction of barred galaxies with significant turnover in \dindex,
\ewhda\ and \ewha\ are 15/32, 13/32 and 14/32, respectively.

We define a galaxy to be a ``turnover'' galaxy if it presents a
significant turnover feature in \dindex, that is, falling below the
lower $1-\sigma$ line in the left-most panel of
Figure~\ref{fig:obs_vs_extra}. Out of the 57 galaxies in our sample,
17 are classified as ``turnover''. These galaxies are highlighted as
big circles in every panel of the figure. We see that, most of the
turnover galaxies defined by the \dindex\ also show turnover feature
in the other two parameters. Table \ref{tbl2} lists the observed and
extrapolated values of the central \dindex, as well as the slope of
the inner \dindex\ profile derived from the linear fit.  Table
\ref{tbl3} lists the number of turnover galaxies identified according
to D$_{n}$(4000), EW(H$\delta_{A}$) and EW(H$\alpha$) profiles.

A striking result that one can immediately identify from this figure
is that, all but two of the turnover galaxies have a bar structure.
Considering that the turnover is observed in only about half of the
barred galaxies, our finding suggests that the presence of a bar is a
necessary, but not sufficient condition for a galaxy to show the
turnover feature. The fact that most of our turnover galaxies
consistently show the turnover feature in all the three parameters can
be regarded as clear signature of recent and ongoing star formation in
their central region. Therefore, our results provide direct evidence
for the central star formation as induced by bar-driven instabilities.

It is interesting to note that, the turnover galaxies are dominated by
those galaxies which have intermediate {\it observed} values of the
diagnostic parameters. In consistent with this, the barred galaxies
without a turnover feature show bi-modal distributions in these
parameters. For instance, they can be well separated into two groups
according to \dindex: one with the largest \dindex\ (\dindex$>1.8$,
thus dominated by old stellar populations), and one with very small
\dindex\ (\dindex$<1.6$, thus indicative of stellar populations
younger than 1-2 Gyr). Such bi-modal distribution gives rise to a gap
in a given parameter, both observed and extrapolated, for the barred
galaxies with no turnover. Our result implies that, the barred
galaxies with a central bulge of intermediate stellar ages are more
unstable to bar-driven instabilities, when compared to those with
either old or young bulges.

\begin{figure*}[t!]
\centering 
\includegraphics[width=0.3\textwidth]{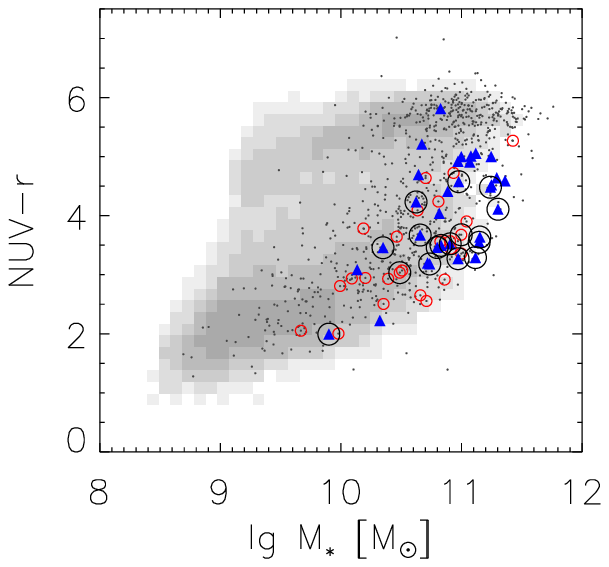}
\includegraphics[width=0.3\textwidth]{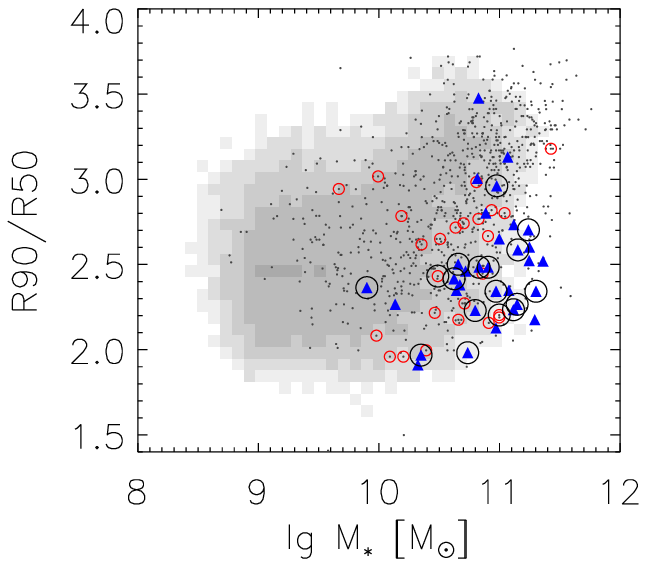}
\includegraphics[width=0.3\textwidth]{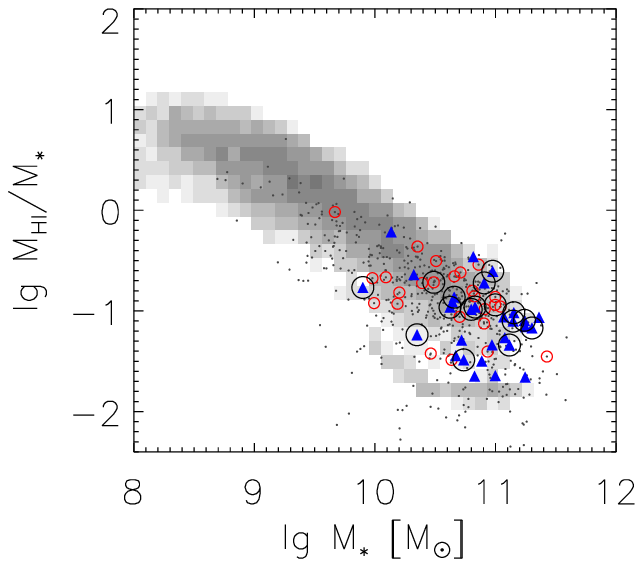}
\caption{The 57 face-on spiral galaxies studied in this work are
  plotted as red circles (unbarred galaxies) or blue triangles (barred
  galaxies), on the \nuvr-\lgmstar\ plane (left panel), 
  \concentration-\lgmstar\ plane (middle panel), and \hifrac-\lgmstar\
  plane (right panel). The galaxies with
  significant turnover in the central \dindex\ (see text for details)
  are highlighted as large black circles. Plotted in small dots are
  the 939 galaxies in the CALIFA mother sample. Distributions of a
  volume-limited sample selected from the SDSS are shown as gray scale
  for comparison.}
\label{fig:color_mass}
\end{figure*}

\begin{figure*}[t!]
\centering
\epsfig{figure=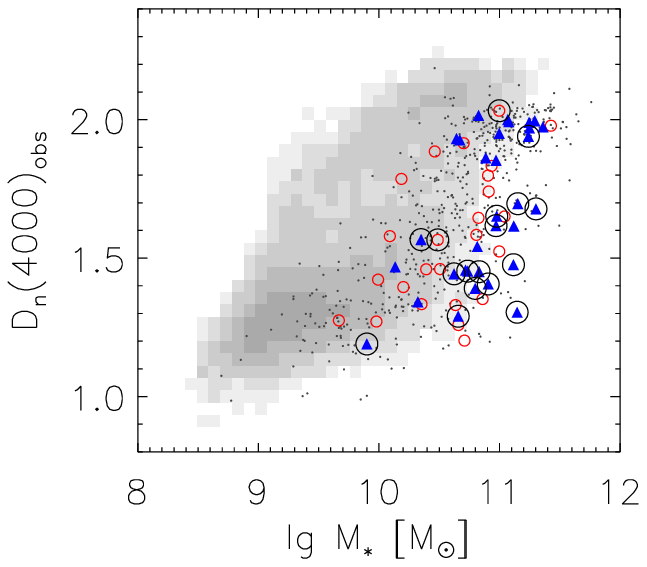,clip=true,width=0.33\textwidth}
\epsfig{figure=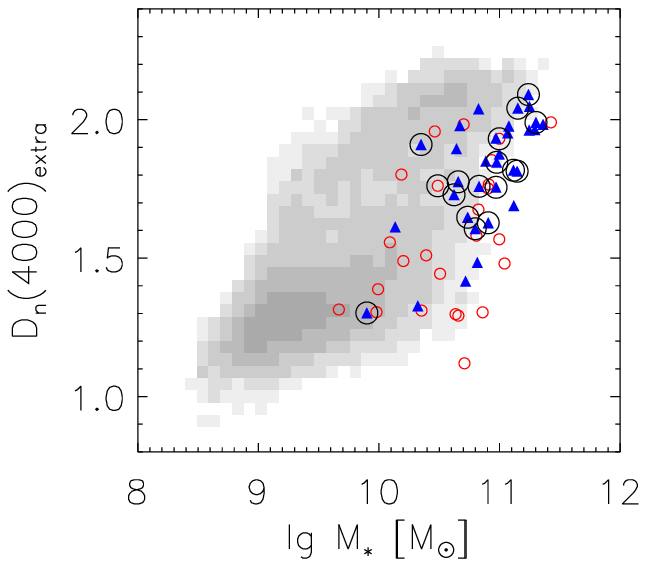,clip=true,width=0.33\textwidth}
\caption{The 57 face-on spiral galaxies studied in this work are
  plotted on the diagram of central \dindex\ versus stellar mass, with
  the same symbols/colors as in the previous figure. For the galaxies
  with a \dindex\ turnover feature, the left (right) panel plots the
  observed (extrapolated) value of the central \dindex\ (see text for
  details). The grayscale maps and small dots present the SDSS sample
  and the CALIFA mother sample, for comparison.}
\label{fig:dindex_mass}
\end{figure*}

\begin{figure*}[t!]
\centering
\epsfig{figure=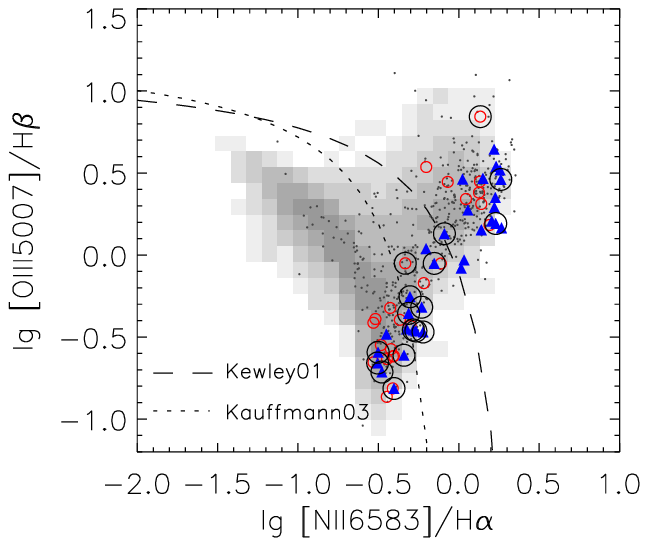,height=0.33\textwidth}
\epsfig{figure=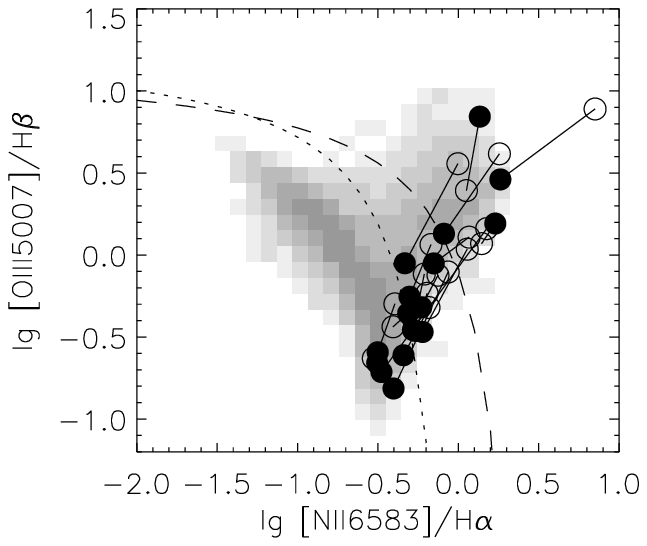,height=0.33\textwidth}
\caption{In the left panel the 57 face-on spiral galaxies studied in
  this  paper are plotted on the BPT \citep{baldwin81} diagram, using
  the same symbols/colors as in Figure~\ref{fig:color_mass}. The
  relevant emission line fluxes are measured from the central spaxel
  of the CALIFA IFU. The right panel presents the BPT diagram for the
  turnover galaxies only, but showing the results for both the
  observed emission line fluxes (filled circles),  and the fluxes from
  inward extrapolating the linear fit of the emission line  flux
  profiles (open circles). For each galaxy the observed and
  extrapolated values are connected by a solid line. Distributions of
  a volume-limited sample of galaxies selected from SDSS are shown as
  grayscale maps in both panels, and 939 galaxies from the CALIFA
  mother sample are plotted small dots in the left panel, for
  comparison. }
\label{fig:bpt}
\end{figure*}

\subsection{Global properties of the turnover galaxies} \label{sec3.2.1}

In this subsection we will examine the global properties of the turnover
galaxies, including their stellar mass (\mstar), NUV-to-optical color (\nuvr),
concentration index (\concentration), \hi\ gas fraction (\hifrac) and nuclear
activity. We take estimates of \mstar, \nuvr\ and \concentration\ from the {\tt
version v0\_1\_2} catalog of NASA-Sloan Atlas (NSA) \footnote{NSA:
http://www.nsatlas.org}, which is a catalog of images and parameters of local
galaxies based on data from SDSS \citep{york00}, GALEX \citep{martin05} and
2MASS \citep{skrutskie06}; see \citet{blanton05a,blanton05b,blanton11} for
details.  The stellar mass of each galaxy is estimated by \citet{blanton07}
from the SDSS five-band Petrosian magnitudes. \nuvr\ is given by the integrated
light in the NUV band from GALEX and the $r$-band Petrosian magnitude from
SDSS. Both are corrected for Galactic extinction. The concentration index is
defined as the ratio of $R_{90}$ to $R_{50}$, the radii enclosing 90\% and 50\%
of the total light in the SDSS $r$-band image.  Furthermore, the \hi\ gas
masses have been collected from HyperLeda database \citep{makarov14}
\footnote{HyperLeda: http://leda.univ-lyon1.fr/}.  A part of galaxies are
covered by ALFALFA survey (Haynes et al., in preparation; results for 70\%
coverage) \footnote{ALFALFA: http://egg.astro.cornell.edu/alfalfa/}.  Since the
errors of \hi\ fluxes from ALFALFA are smaller, we replaced the ALFALFA \hi\
measurements for the overlapped sources. 
 
Figure \ref{fig:color_mass} displays the global properties of our galaxies:
\nuvr\ versus \lgmstar, \concentration\ versus \lgmstar, and \hifrac\ versus
\lgmstar, highlighting the barred, unbarred and turnover galaxies with the same
symbols/colors as in the previous figure. For comparison, we have selected a
volume-limited sample of 35070 galaxies with stellar mass above
10$^{8}$M$_\sun$ and redshift in the range $0.01<z<0.03$ from the NSA.
Distributions of this sample are plotted in grayscale maps in both panels.
When compared to the SDSS sample, the CALIFA mother sample lacks galaxies of
low mass, particularly for the red population. However, as can be seen from the
figure, the CALIFA sample is still representative of the general population at
masses above $\sim10^{10}$M$_\sun$ (see \citealt{walcher14} for a more
quantitative demonstration). As expected, the CALIFA galaxies are well
separated into the red sequence with \nuvr$>5$ and the blue cloud with
\nuvr$<4$, with a minority of the sample falling in the green valley in
between.

Most of the galaxies in our sample are located in the blue cloud and
green valley in the color-mass diagram, and have concentration indices
smaller than \concentration$=2.6$, indicative of late-type
morphologies.  This is expected as our sample is limited to spiral
galaxies by selection. The unbarred galaxies in our sample are
preferentially found in the blue cloud, with only a few extending to
the green valley. In contrast, the barred galaxies are distributed
over both the green valley and blue cloud regions. Looking at the
barred galaxies only, we find the galaxies with and without turnover
feature to show quite distinct distributions in the color-mass diagram
--- the non-turnover galaxies mostly fall in the green valley, but the
turnover population is biased towards the blue cloud. The same
trends/differences are not seen in the
\concentration-\lgmstar\ diagram, where the different types of
galaxies appear to occupy similar regions.

In Figure~\ref{fig:dindex_mass} we present the
\dindex-\lgmstar\ diagram which plots the \dindex\ as measured from
the central region of the galaxies as a function of their total
stellar mass.  For the background SDSS sample and the CALIFA mother
sample, we take the \dindex\ measurements from the NSA which are based
on the SDSS 3\arcsec-fiber spectra. Since SDSS spectroscopy is
available only for half of the selected 57 face-on spiral galaxies,
for these galaxies we instead plot the \dindex\ measured from the
central spaxel of their CALIFA datacubes. We note that the CALIFA
samples, both the mother sample and the subset studied here, appear to
lack galaxies with highest \dindex\ compared to the SDSS sample,
although the CALIFA samples do include galaxies of highest
\nuvr\ colors. We have compared the \dindex\ as obtained from the SDSS
and the CALIFA central spaxel for a sample of CALIFA/DR2 galaxies with
data available from both surveys. The two measurements agree well, but
the IFU central spaxel gives a value that is slightly larger than the
value from the SDSS 3\arcsec-fiber, with difference of
$\Delta$\dindex$\sim0.05$. This means the lack of galaxies with
highest \dindex\ is indeed the case for the CALIFA samples, and this
cannot be attributed to this slight difference between the CALIFA and
SDSS measurements. The slight difference may be a combined effect of
the possible errors in fiber centroiding and seeing/resolution effect
\citep{sanchez12}, as well as the negative gradient of the
\dindex\ profiles of galaxies. Nevertheless, this small effect is
neglectable for the comparisons of the different samples presented
here.

Similar to the \nuvr-\lgmstar\ diagram, the turnover galaxies are
observed to be located in the young population regions which can be
defined by \dindex$<1.6$, while the barred galaxies without turnover
feature mostly have \dindex$>1.6$. When plotted using the extrapolated
\dindex\ (the middle panel), the turnover galaxies are located mostly
above the empirical divider of \dindex$=1.6$. As a result, the barred
galaxies as a whole are largely confined to the area of old
populations, with only a few galaxies scattering towards the young
population region. A reasonable conjuncture based on our result is
that, most barred galaxies in our sample had their central star
formation quenched at an earlier time, but for some reason some of
these galaxies (the turnover galaxies) recently experienced, or is
experiencing a new round of star formation, thus moving from the old
population region back to the young population region in the
\dindex-\lgmstar\ diagram.

\begin{figure*}[t!]
\centering \includegraphics[width=\textwidth]{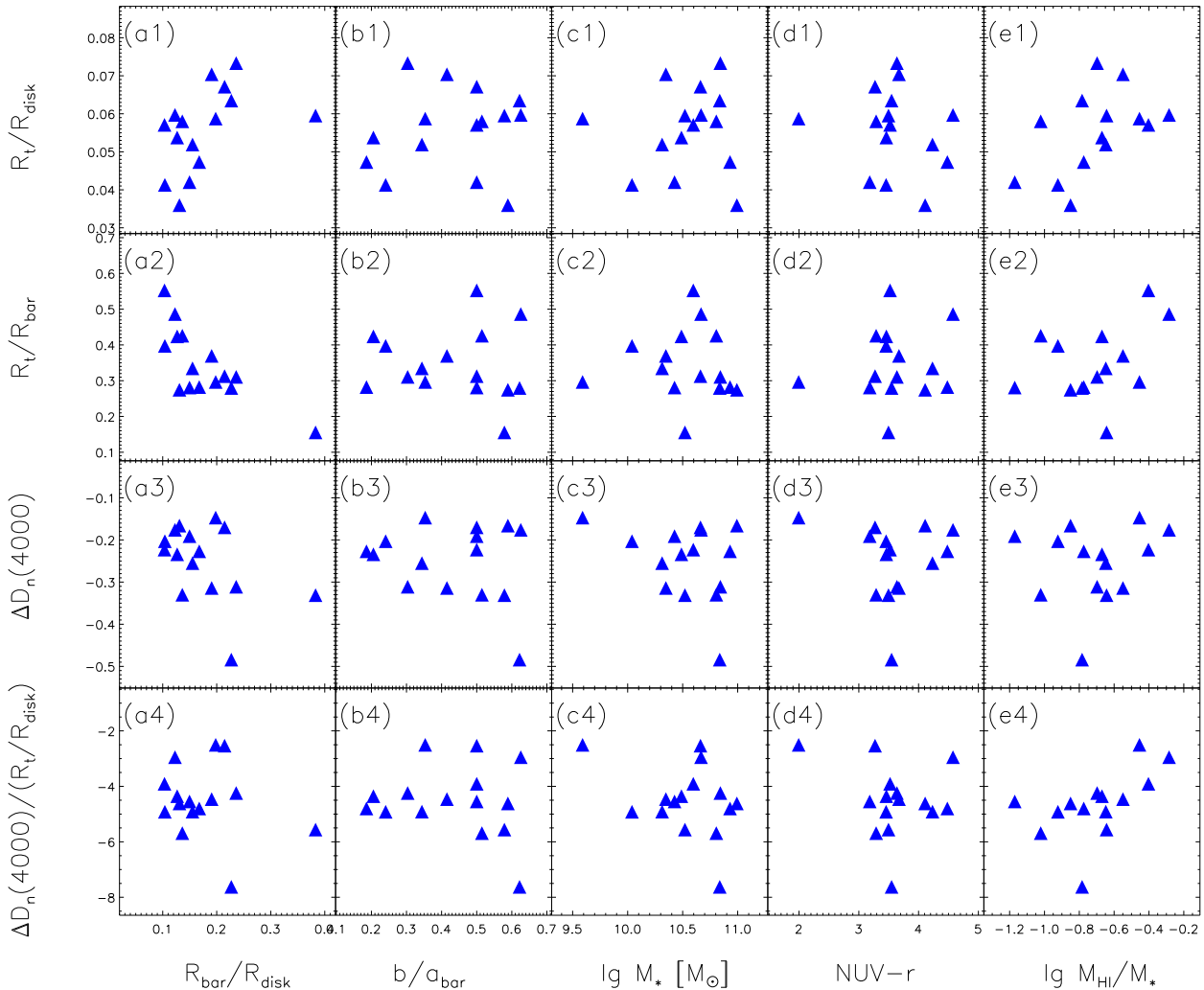}
\caption{The size and strength of the turnover region  are plotted as
  a function of galaxy properties. The y-axis of  panels from top to
  bottom shows the turnover-to-disk radius ratio (\rt/\rdisk), the
  turnover-to-bar radius ratio (\rt/\rbar), the drop in central
  \dindex\ ($\Delta$\dindex) and the  turnover strength-to-size ratio
  ($\Delta$\dindex/(\rt/\rdisk)).  The x-axis of panels from left to
  right is the bar-to-disk radius ratio (\rbar/\rdisk), the
  minor-to-major axis ratio of the bar ($b/a_{\mbox bar}$), stellar
  mass (\lgmstar) and the \nuvr\ color.}
\label{fig:scaling_relations}
\end{figure*}

Figure~\ref{fig:bpt} presents our samples on the BPT diagram
\citep{baldwin81}.  The fluxes of the four relevant emission lines are
measured either from the SDSS 3\arcsec-fiber spectra for the SDSS
sample, or from the central spaxel of the CALIFA datacubes for both
the CALIFA mother sample and the 57 galaxies studied here. As can be
seen from the left panel, our sample cover a wide area in the BPT
diagram, including star-forming, AGN/SF composite, and AGN host
galaxies, although the sample lacks both Seyferts and strongly
star-forming galaxies due to the lack of low-mass galaxies in the
CALIFA survey. The distributions of barred and unbarred galaxies are
similar, consistent with previous findings that the presence of a
prominent bar is not closely linked to the nuclear activity of
galaxies \citep[e.g.][]{lee12,cheung15}. It is interesting that,
although the barred galaxies as a whole are distributed over all the
three regions on the BPT diagram, those falling in the AGN region are
mostly non-turnover galaxies, while the turnover galaxies are
predominantly found in the star-forming and composite regions. Among
the 17 turnover galaxies, the number of galaxies in the
AGN/composite/SF regions are 4, 8, 5, respectively. For the 15
non-turnover barred galaxies, only two are located outside the AGN
region, with one composite and one SF galaxy.

The distinct behavior of the turnover and non-turnover galaxies on the
BPT diagram is not surprising, if the turnover in the SFH diagnostic
parameters is a result of the central star formation caused by the
bar-driven gas inflow. In this case, the emission line fluxes of the
central region used for the BPT diagram may also deviate from their
normal values, thus leading the galaxy to move to a different location
on the diagram. To test this hypothesis, we have measured the profiles
for the four emission lines and quantify the central turnover feature
in the same way as we did for the three SFH diagnostic parameters.  In
the right-hand panel of Figure~\ref{fig:bpt} we plot the turnover
galaxies, i.e. those identified according to the \dindex\ profile,
using both the observed emission line fluxes (filled circles) and the
fluxes obtained from extrapolating the linear fit profile (open
circles). All the galaxies move their location, though to varying
degrees. This confirms that the relevant emission lines also show
turnover feature in the galactic center, like what we find for the SFH
diagnostic parameters, and bar-driven gas inflow can well be the
physical process driving all these turnover features. 

\subsection{Correlations of the turnover feature with host galaxy properties}

In the previous subsection we have examined the global properties of
the turnover galaxies. In this subsection we will characterize the
turnover feature and examine the correlation with host galaxy
properties.  

For each turnover galaxy, we estimate two quantities, \ddindex\ and
\rt, to respectively characterize the strength and size of the
turnover feature. Here, \ddindex\ is defined as the difference of the
observed \dindex\ in the central spaxel relative to the central
\dindex\ obtained from extrapolating the linear fit.  The radius of
the turnover region, \rt, is determined from the \dindex\ profile of a
given galaxy, defined to be the maximum radius along major axis beyond
which the observed \dindex\ no longer deviates significantly from the
linear fit (see \S\ref{sec:turnover}). We also determine the radius
for both the disk component (\rdisk) and (when applicable) the bar
component (\rbar), the radius along major axis where the disk or bar
component of the surface brightness profile drops to 25 mag
arcsec$^{-2}$ (see \S\ref{sec:photometry}). We use \rtrdisk\ and
\rtrbar\ to describe the relative size of the turnover with respect to
the disk and bar components, and \ddindex/\rt\ to indicate the
specific turnover strength per unit radius. In
Figure~\ref{fig:scaling_relations} we show the turnover properties
including turnover size relative to both disk and bar sizes (\rtrdisk,
\rtrbar), turnover strength (\ddindex) and strength-to-size ratio
(\ddindex/(\rtrdisk)), as a function of host galaxy properties
including bar-to-disk radius ratio (\rbar/\rdisk), minor-to-major axis
ratio of the bar ($b/a_{\mbox{bar}}$), stellar mass (\lgmstar),
\nuvr\ color and \hi\ gas fraction (\hifrac).

Both the turnover size and its strength are not correlated with galaxy mass or
color. And there are no clear correlations with the bar ellipticity as
quantified by the minor-to-major axis ratio of the bar. The probable weak
correlations can be only seen from bar length (the first column $a$) and \hi\
gas fraction (the last column $e$).  As seen in panel $a1$, \rtrdisk\ is
positively correlated with \rbar/\rdisk, that is, galaxies with longer bars
tend to have larger area of turnover feature in their central region when
compared to the galaxies with similar disk size but shorter bars. A similar
correlation between turnover size and disk size can be seen from panel $a2$,
although the correlation is mainly contributed by a few galaxies with highest
\rt/\rbar\ but lowest \rbar/\rdisk.  This might be suggesting that, although
the bar size is more closely linked with the turnover size, for galaxies with
similar sizes of bars, they may also have larger turnover area if their disks
are larger.  As for the \hi\ gas fraction, both \rtrdisk\ and
\ddindex/(\rtrdisk) are larger as the increasing of \hi\ gas fraction. It
suggests that the turnover size and its strength correlate with the amount of
gas in the disk. If there is enough gas supply, the central star formation can
be supported for long timescale.  Unfortunately, we did not find any CO
measurements of our galaxies from the current databases. The direct
observational evidences are still needed to confirm this.

\begin{figure*}[t!]
\centering
\includegraphics[width=0.33\textwidth]{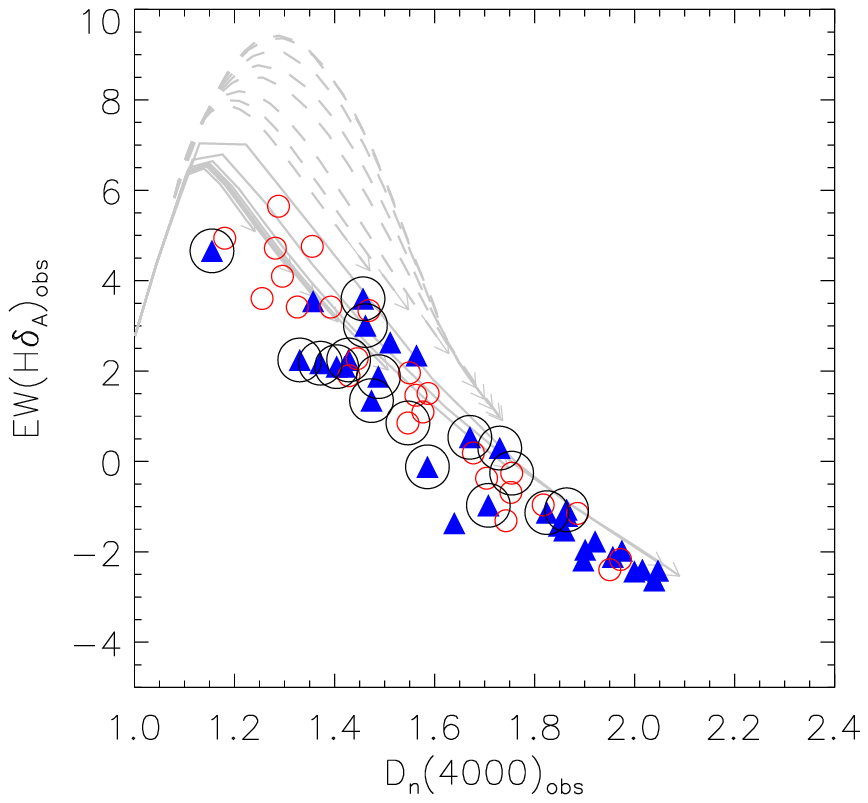}
\includegraphics[width=0.33\textwidth]{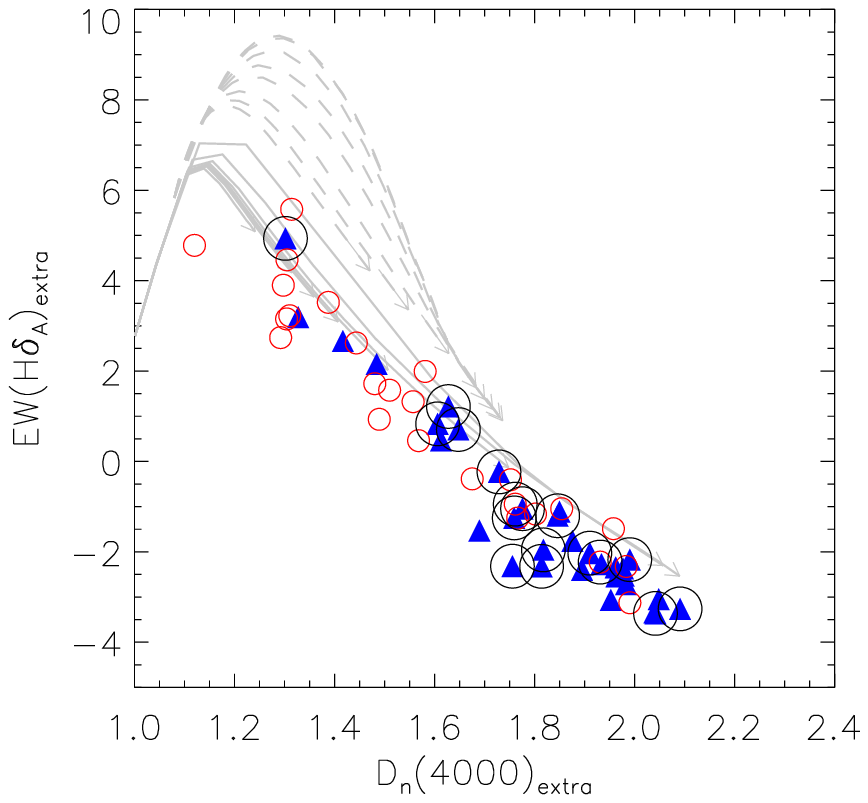}
\includegraphics[width=0.33\textwidth]{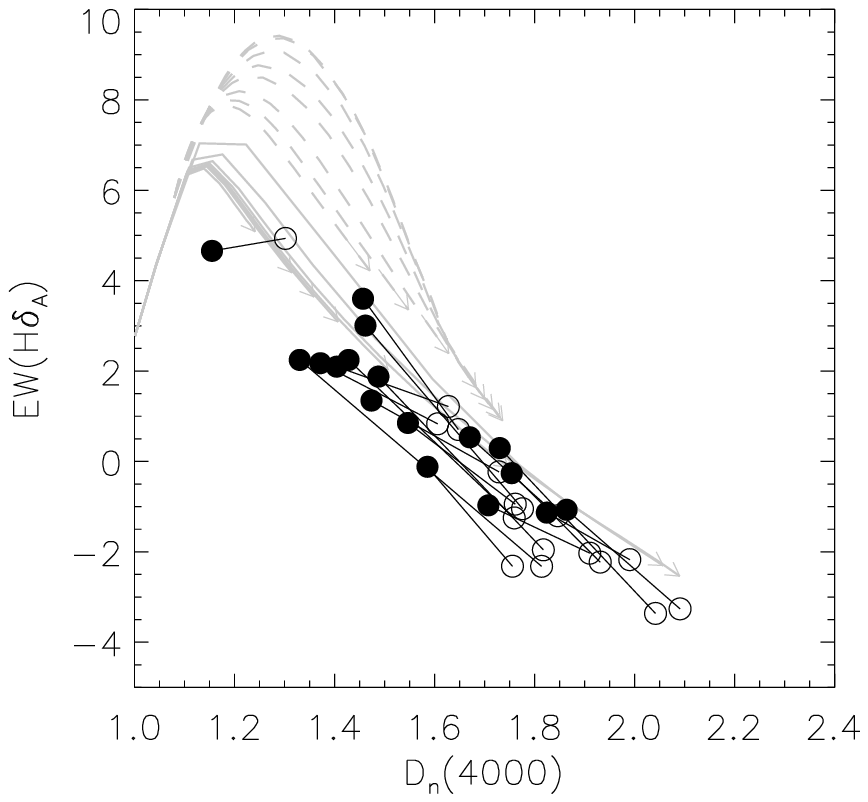}
\includegraphics[width=0.33\textwidth]{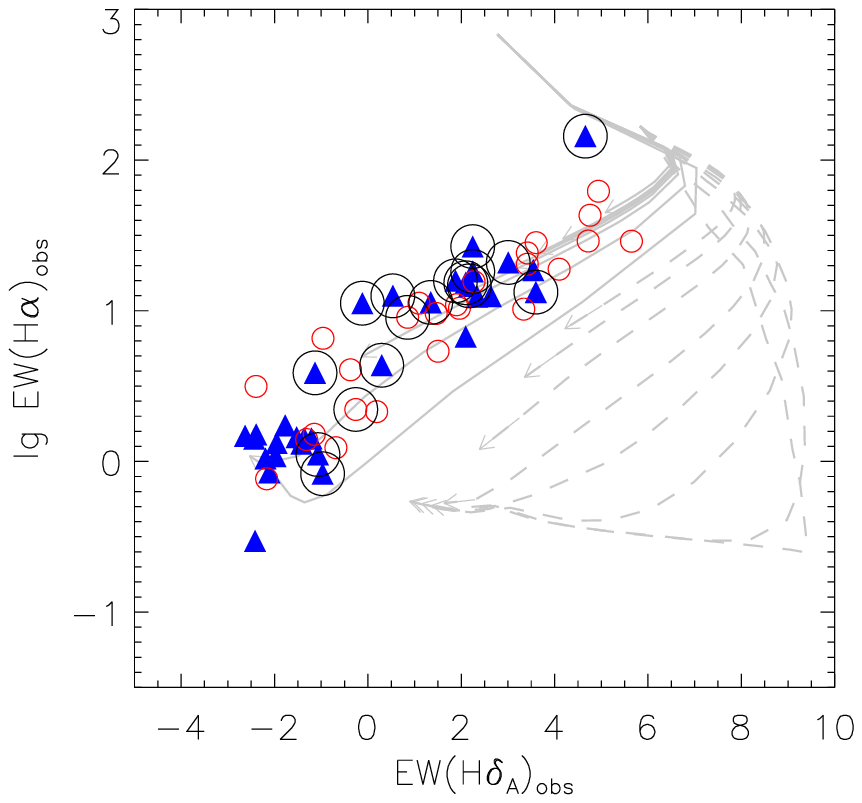}
\includegraphics[width=0.33\textwidth]{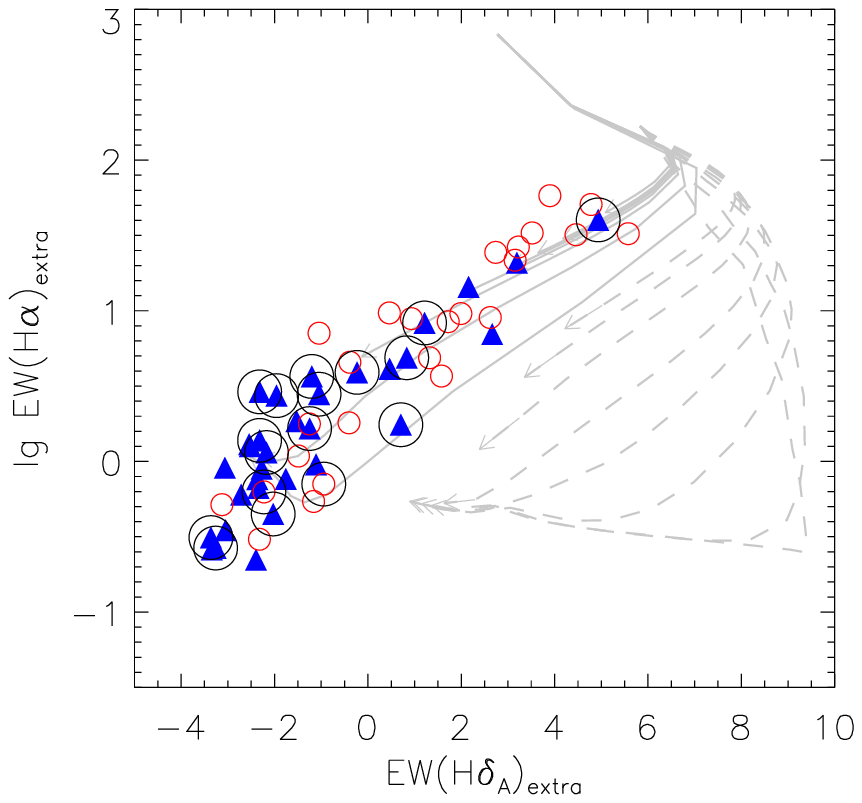}
\includegraphics[width=0.33\textwidth]{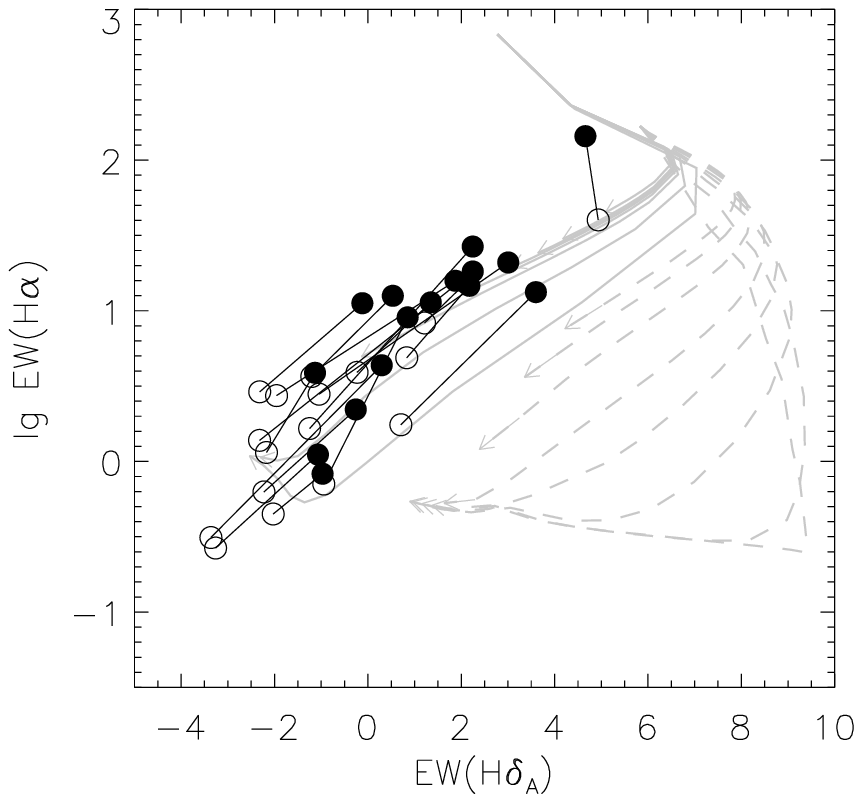}
\caption{Our galaxies are plotted as colorful symbols on planes of
  \dindex\ vs. \ewhda\ (upper panels) and \ewhda\ vs. \ewha\ (lower
  panels). Shown by solid and dashed lines are solar-metallicity
  models of continuous star formation histories and star formation
  bursts from the stellar population synthesis code of
  \cite{bruzual03}.  The left panels are based on the observed
  emission line measurements, and the middle panels based on the
  inward extrapolation of the radial profiles. The right panel shows
  the turnover galaxies only, but for both observed (open circles) and
  inward extrapolated (solid circles) measurements, with the solid
  lines connecting the two measurements of each galaxy.  }
\label{fig:diagnostic_diagrams}
\end{figure*}

\begin{figure*}[t!]
\centering \includegraphics[width=0.8\textwidth]{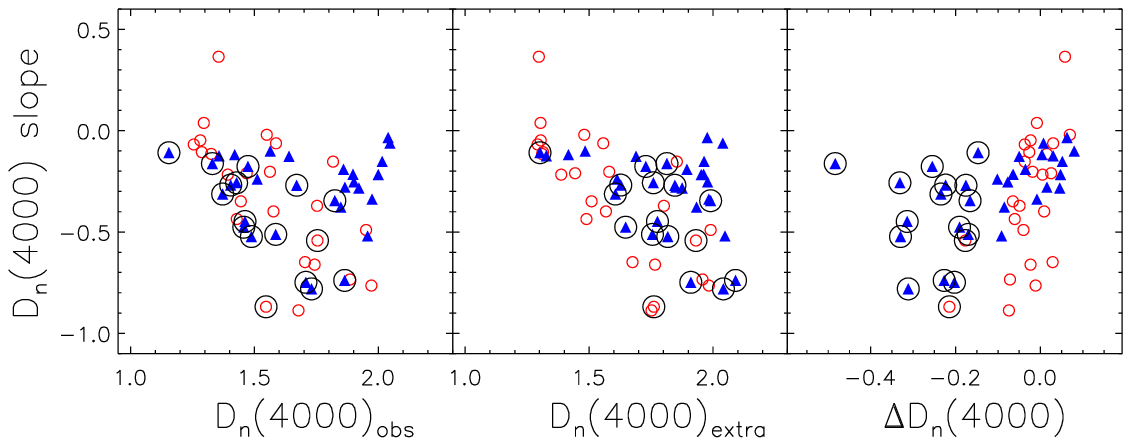}
\includegraphics[width=0.8\textwidth]{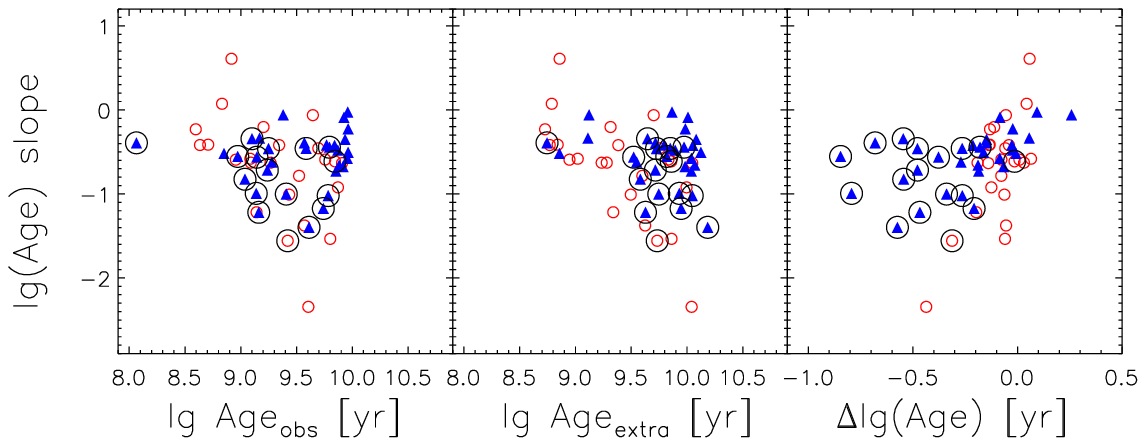}
\caption{{\it Upper:} slope of the inner profile of \dindex\  as a
  function of the observed central \dindex\ (left), the inward
  extrapolation of the central \dindex\ (middle), and the difference
  between the observed and extrapolated central
  \dindex\ (right). Lower: Luminosity-weighted stellar population age
  gradients as a function of the observed central ages (left),
  extrapolated central ages (middle) and $\Delta$age values (right).
  Symbols and colors are the same as in Figure
  \ref{fig:obs_vs_extra}.}
\label{fig:slope_vs_center}
\end{figure*}

\subsection{Recent star formation histories of the turnover galaxies}

In Figure~\ref{fig:diagnostic_diagrams} we display our galaxies on the
diagnostic diagrams of recent star formation history, namely the
planes of \dindex\ versus \ewhda\ (upper panels) and \ewhda\ versus
\ewha\ (lower panels). In the left two columns we show all the
galaxies  in our sample, but for the turnover galaxies, we use the
observed parameters in the first column and the extrapolated
parameters in the second. The barred, unbarred and turnover galaxies
are plotted with blue triangles, red small circles and black big
circles. In the right-most column we display the turnover galaxies
only, showing both the observed and extrapolated values which are
connected by a solid line for a given galaxy. For comparison and in
each panel, we present the same parameters for the solar-metallicity
models of BC03 that follow exponentially declining star formation
histories, with solid lines for continuous star formation decline with
long $e$-folding times ($\tau>5\times10^8$yr) and dashed lines for
bursts of star formation with fairly short $e$-folding time
($\tau<5\times10^8$yr). The \dindex\ and \ewhda\ indices are measured
from the BC03 model spectra adopting the same definitions as the
observed spectra. For \ewha, we compute a H$\alpha$ luminosity from
the output of each BC03 model, by converting Lyman continuum photons
to H$\alpha$ photons following \citet{hunter04} (see Equations
(B2)-(B4) in their Appendix) and adopting the recombination
coefficients and H$\alpha$-to-H$\beta$ ratios from
\citet{hummer87}. We assume nebular temperatures of $T_e=$15,000,
10,000, and 5000K for the LMC, solar-metallicity, and
supersolar-metallicity models, respectively.

Overall, our sample occupies roughly the same region as the models of
continuum star formation decline in both \dindex-\ewhda\ and
\ewhda-\ewha\ planes, with no galaxies falling in the region of star
formation bursts. The unbarred sample covers full ranges in all the
three parameters. When plotted with the observed parameters, the
barred sample lacks galaxies with lowest \dindex, highest \ewhda, and
highest \ewha. When the turnover galaxies  are plotted using the
extrapolated instead of the observed parameters, the barred galaxies
are mostly located in the regions of low-\dindex, high-\ewhda\ and
high-\ewha, consistent with models which contain no/little young
populations and haven't experienced star formation in the past 1-2
Gyr. The difference between the first and second columns is caused by
the turnover galaxies which are observed to be located in the
intermediate regions and move to the low-\dindex\ and
high-\ewhda\ region when plotted with the extrapolated values. The
systematic change of their locations is seen more clearly in the
right-most panels which show the turnover galaxies only, but using
both the observed and the extrapolated values.

In Figure~\ref{fig:slope_vs_center} (upper panels), we plot the slope
of the \dindex\ profile obtained in \S\ref{sec:maps} by fitting a
linear function to the inner region, as a function of the observed
(left) and extrapolated (middle) central \dindex\ (middle), and the
difference between the two indices (right). All the galaxies except
one present a negative or flat slope. For unbarred galaxies (red open
circles), the \dindex\ slope shows good correlations with both the
observed and the extrapolated \dindex, in the sense that galaxies with
a larger \dindex\ in their center present steeper (more negative)
slope, namely, having smaller \dindex\ in the outskirt compared to the
central region. The \dindex\ slope increases from around zero at the
smallest central \dindex\ ($\sim1.2$) to $\sim-0.8$ at the largest
central \dindex\ ($\sim2$). This correlation is consistent with the
{\em inside-out} picture of the star formation quenching in galaxies,
in which the star formation gets shut down firstly in the galactic
center, before extending to larger and larger radii.

Different to the unbarred galaxies, the barred galaxies as a whole
show no clear correlations in either case. It is interesting that,
however, they behave quite differently when divided into two subsets,
with and without turnover feature. The barred galaxies without
turnover feature are mostly located in the region with
high-\dindex\ and flat slope, well deviating from the area of the
unbarred galaxies.  The large central \dindex\ and the flat slope
combine to suggest that both the center and the outer region of these
galaxies are dominated by old stellar populations. In contrast, the
turnover galaxies follow a tight correlation between the observed
central \dindex\ and the slope, which is similar to, but even tighter
than the unbarred galaxies. When plotted with the extrapolated
\dindex\, the turnover galaxies move towards area of larger \dindex,
located in between the unbarred galaxies and the non-turnover barred
galaxies.  The different behaviors of the different types of galaxies
are more clearly shown in the right-most panel of the same figure
where the \dindex\ slope is plotted against the
observed-to-extrapolated  \dindex\ difference. 

As a complementary analysis, the lower panel of
Figure~\ref{fig:slope_vs_center} shows the gradient of luminosity-weighted
stellar age as a function of the observed (left) and extrapolated
(middle) central age, and the difference between the two. The stellar
age of given spaxel in the datacubes is estimated using the {\tt
  STARLIGHT} code as described in \S\ref{sec:data}. The age gradient
is given by the slope of the linear fit to the inner region of the age
profile, over the same radial range adopted to fit the
\dindex\ profile. The central extrapolated age is obtained by
extrapolating the fitted profile towards the galactic center, in the
same ways as in case of \dindex. The behaviors of the central ages and
age gradient are in good agreement with what are shown in the upper
panels.  We have also estimated mass-weighted ages, finding much
smaller gradients and weaker turnover features when compared to those
from the luminosity-weighted ages. This can be understood from the
fact that young stellar populations contribute a large fraction of the
total luminosity but the stellar mass is more dominated by old
populations.  Therefore, the flatter profile of the mass-weighted
profiles and the lack of central turnover suggest that, it is the
young populations formed in the recent past that dominate both the
central turnover feature and the overall stellar population gradient.

\section{Discussion}
\label{sec:discussion}

\subsection{Is the turnover always present simultaneously in the three 
parameters?}
\label{sec:individuals}

In this work we have defined our turnover galaxies according to their
\dindex\ profile only. We note that, 5 of the 57 galaxies in our
sample  present significant central upturn in \ewhda\ or \ewha, but
they show no turnover feature in \dindex. We divide them into two
subsets and discuss their properties separately in the following two
paragraphs.

The first subset consists of 4 galaxies: {\sl IC~1256}, {\sl
  NGC~3106}, {\sl UGC~11649} and {\sl UGC~12224}. These galaxies
present a central upturn in \ewha, with no significant turnover
feature in \dindex\ or \ewhda. This subset includes both barred and
unbarred galaxies (see Table~\ref{tbl1}), with their center being
classified as either star-forming region or LINER on the BPT
diagram. {\sl NGC~3106} shows ring-like tidal tails in the outskirt,
with a close faint companion galaxy in the SDSS image. {\sl UGC~11649}
is a barred galaxy, with an outer ring-like structure.  {\sl IC~1256}
is an unbarred spiral with an inner ring, with surrounding faint
galaxies shown in the SDSS image. We note that this galaxy also show
turnover in \ewhda, thought not in \dindex.  {\sl UGC~12224} is a Sc
type galaxy with very diffuse spiral arms. The upturn of
\ewha\ indicates the enhancement of central star formation in these
galaxies, which might be triggered by tidal interactions (e.g. {\sl
  NGC~3106}, {\sl IC~1256}) or other mechanisms. The central star
formation enhancement must have started very recently, considering the
lack of similar turnover feature in \dindex\ and \ewhda\ which
indicate stellar populations formed in the past.

The other subset has only one galaxy, {\sl NGC~7563}, which shows
turnover only in \ewhda. This is a barred, early-type (Sa) spiral.  It
has a faint ring-like structure in the outskirt, with no obvious
spiral arms. Color-mass diagram suggests this is a passive red
galaxy. The \dindex\ is above 1.6 across the galaxy, and the
\ha\ emission is very weak everywhere. The enhanced \ewhda\ in the
center implies the existence of massive young stars, and thus
rejuvenate recent star formation. The origin of this star formation,
however, is unclear  based on our data alone.

Having discussed these exceptional cases where \dindex\ turnover  is
not associated with the turnover in \ewhda\ or \ewha, now we come back
to the galaxies with \dindex\ turnover. As mentioned earlier
(\S\ref{sec:turnover}), most of the galaxies with turnover in
\dindex\ also present turnover feature in \ewhda\ and/or
\ewha. Indeed, this is the case for 12 out of 15 galaxies that we
identify  to have a central turnover in \dindex.

The rest three galaxies are {\sl NGC~0036}, {\sl NGC~3381} and {\sl
  NGC~4210}.  They show turnover feature in \dindex, but not in the
other two parameters. We find that, the \dindex\ turnover in {\sl
  NGC~0036} and {\sl NGC~4210} are actually very weak; the profiles of
all the three parameters are close to be flat, with slight decrease in
\dindex\ and weak upturn in \ewhda\ and \ewha\ towards the smallest
radii. In Figure~\ref{fig:obs_vs_extra}, the two galaxies are very
close to the $1-\sigma$ line which separates the turnover and
non-turnover classes. The third galaxy, {\sl NGC~3381}, similarly
shows a weak turnover feature in both \dindex\ and \ewha, but there is
no obvious turnover in \ewhda. Therefore, for these galaxies, the
central behaviors of the three parameters are not  necessarily
inconsistent with each other, and their different turnover classes
should not be emphasized. We should be safe to claim that, most (if
not all) of the galaxies with a \dindex\ turnover also present
turnover feature in both \ewhda\ and \ewha.

This finding is striking given the known fact that the three
diagnostic parameters are sensitive to different ages of stellar
populations.  The fact that the central enhancement in both
\ewhda\ and \ewha\ are always associated with the central decrease of
\dindex\ implies  that, no matter what physical processes are
responsible for the turnover feature, they must have been at work
continuously since at least 1-2 Gyr ago, and are still working at the
present day. It is natural to conjecture that the several galaxies
with central turnover only in \ewha\ but no feature in other
parameters are at the early stage of this long term event, expected to
present turnover feature in \ewhda\ and \dindex\ at later times. We
would like to point that, however, in order to better understand these
exceptional galaxies, one would need larger samples that allow better
statistics.

\subsection{Is the presence of a bar a necessary condition for the 
turnover feature?}

The most striking result produced by the current work is that, almost
all the turnover galaxies in our sample host a bar structure, whereas
only half of the barred galaxies present turnover feature. This result
appears to suggest that the presence  of a bar is a necessary, but not
sufficient condition for the central turnover to occur.

The three diagnostic parameters are known to be sensitive indicators
of different ages of stellar populations.  Therefore, for a given
spaxel in the datacubes, these parameters combine to roughly tell the
recent star formation history of the corresponding region within the
galaxy. The ``turnover'' of these diagnostic parameters thus reveals
recent/ongoing star formation occurring in the galactic center over
the past 1-2 Gyr, which makes the central region younger than one
would expect for a galaxy of the same kind. Our work thus provides
direct evidence for the bar-induced star formation enhancement in the
central region of galaxies. 

The contrast in nuclear star formation between barred and unbarred
galaxies was first observed more than forty years ago based on radio
continuum surveys
\citep{cameron71,dressel78,dressel79,heckman80,hummel80,hummel81}. By
contrasting the global radio luminosity with the nuclear radio
luminosity for both barred and unbarred galaxies, these early studies
detected a marked excess of nuclear star formation in $\sim30-50$\% of
barred systems (see reviews by
\citet{kennicutt94,kennicutt98,kormendy04}). In a recent work,
\citet{wang12} analyzed a sample of $\sim$1500 barred galaxies
identified from SDSS, finding the central star formation rate (as
inferred by \ha\ luminosity measured from the SDSS 3\arcsec-fiber
spectroscopy) to be enhanced in strongly barred galaxies. These
authors also found that not all the barred galaxies showed central
enhanced star formation rates.  Our results are in good agreement with
these previous studies. With the help of both the resolved
spectroscopy from the CALIFA IFU and the high resolution optical image
from SDSS, we are allowed to probe the radial gradient of the recent
star formation history for different structural components separately.

However, \citet{wang12} found that only 50\% of the galaxies with
central SFR enhancement host a strong bar. In contrast, we find almost
all the turnover galaxies are barred. This discrepancy may be
attributed to the different sample selections adopted in the two
studies.  Although both works select relatively face-on late-type
galaxies, one important difference is that our sample excludes
merging/interacting systems. As those authors suggested, the central
SFR enhancement in the unbarred galaxies may be caused by other
processes such as galaxy-galaxy interactions. In consistent with this
suggestion, a study of a complete sample of $\sim10^{5}$ star-forming
galaxies in SDSS revealed that only $\sim$40\% of the most strongly
star-forming systems have companion galaxies within $\sim70$
kpc. Therefore, based on both \citet{li08} and \citet{wang12}, it is
likely that both the internal bar structure and tidal interactions
with other galaxies are responsible for the central star formation
enhancement as observed in a complete sample of nearby galaxies. It'd
be interesting to extend the IFU-based analysis presented in this
paper to samples of merging galaxies, galaxies with close companions
and those located in crowded regions. This will be our next step, in
which we will make use of the ongoing MaNGA survey which provides much
larger sample covering wider ranges in galaxy mass, color, morphology
and environment.

\subsection{Implications for bar-driven gas inflow as origin of the
central turnover feature}

Both theoretical studies and $N$-body simulations have well
established that bars in galaxies grow through transferring angular
momentum to the outer disk, which drives the gas outward (if located
outside corotation) to form an outer ring, or inward (if inside
corotation) to fall to the center \citep[see][for a
  review]{sellwood93}.  This gas infall is believed to make
pseudobulges through the induced star formation in the central region
\citep[see][and references therein]{kormendy04}.  Our results agree
well with this picture. The color-mass diagram in
Figure~\ref{fig:color_mass} shows that the turnover galaxies are
mostly bluer than \nuvr$\sim4-5$, bluer than those non-turnover barred
galaxies. Since \nuvr\ is a sensitive indicator of the cold gas
content of galaxies \citep{zhang09,li12}, our result suggests  the
turnover galaxies contain more cold gas than those non-turnover
galaxies. This supports the bar-driven gas inflow as the driving
process for the central star formation and the resulting turnover
feature, as observed in some of the barred galaxies. This might be one
of the reasons for the absence of turnover feature in half of our
barred galaxies which have redder \nuvr\ color and thus don't have
substantially high fraction of cold gas in their disk. 

The intermittent nature of bar-induced gas inflow might be another
reason why the turnover is not always seen in barred galaxies.
Simulations show that the inflowing gas will first be accumulated at
the ends of the bar, before it can further fall inward through loss of angular
momentum \citep{combes07}. The inflow timescale is very short ($\sim$10 Myr)
compared with the typical lifetime of bars \citep{combes07}.  Similar
hints come from studies of star-forming nuclear rings which are also
related to the bar-driven gas transport from disk to galactic
center. \citet{sarzi07} found that circumnuclear star formation is
most likely to experience episodic starbursts rather than constant
star formation. High star formation rates in the central part of the
galaxy can exhaust the gas very quickly, turning off the star
formation, which will restart until more gas is supplied
inside. However, our \dindex-\ewhda\ and \ewhda-\ewha\ diagrams show
that the central star formation history in the turnover galaxies
closely follows models of continuous star formation, and this is true
for both the observed and the extrapolated values of the diagnostic
parameters (see Figure~\ref{fig:diagnostic_diagrams}). 

Based on CALIFA/DR1 data, \citet{holmes15} have attempted to search
for  bar-like bisymmetric non-circular flow, believed to be a
signature of  gas inflow. We note that 6 of our turnover galaxies were
included in that work, of which 3 showed non-circular flows ({\sl
  NGC~0477}, {\sl UGC~03253} and {\sl NGC~7321}) and 1 presented
warped flow ({\sl NGC~0036}).  The warped flow may be not driven by
bars, but rather results from an oval disk which drives gas infall in
a similar way to what a bar does (Kormendy \& Kennicutt 2004, and
references therein). Considering the short time scale of gas infall
and the relatively longer period of the induced star formation,
perhaps it is not surprising to see no non-circular flow in some of
our turnover galaxies. 

In addition to bar-driven instabilities and tidal interactions
discussed above, other mechanisms have been suggested to pull more gas
inward from outer disks, such as spiral-driven instabilities
\citep{sellwood11} and angular momentum exchange with the hosting dark
matter halo \citep{athanassoula03}. In any case, a common purpose of
all these mechanisms is to bring the cold gas in disk to the central
region. In order for a complete picture of the driving processes for
the gas inflow, one would need to observationally map the molecular
gas distribution from outer disks down to small galactic centric radii
with spatial resolution comparable to the current IFU instruments.

\subsection{Implications for growth/rejuvenation of pseudobulges}

Our finding of the recent/ongoing star formation as inferred from the
central turnover of the diagnostic parameters provides strong support
to theoretical expectations that bar-driven gas flow triggers star
formation activity in the central region, which leads to ongoing
growth of the central (pseudo) bulge. In fact, the role of bars in
building up pseudobulges in galactic centers, and consequently driving
the secular evolution of galaxies, has a rich history of research over
more than four decades, and obtained great attention particularly in
the past decade or so \citep[see the comprehensive reviews
  by][]{kormendy93, sellwood93,kormendy04,sellwood14,kormendy13}. As
reviewed by \citet{kormendy04}, many barred galaxies (and oval
galaxies) have very active star formation in their bulge. In a recent
work, \citet{coelho11} applied stellar population synthesis to a
sample of 575 galaxy bulges in the SDSS, finding the barred galaxies
with \lgmstar$\ga10$ to show a bimodel distribution of bulge ages: a
normal distribution with mean age of 10.7 Gyr which is comparable to
unbarred galaxies of similar stellar masses, and an excessive young
population with mean age of 4.7 Gyr which is not seen in unbarred
galaxies. The authors suggested that the excess of young population in
the bulges was an evidence for the rejuvenation of bulges as induced
by bar-driven star formation. Simulation analysis by \citet{spinoso16} 
also shown evidence that bar has the impact on the evolution of the 
gas and stellar components of the galaxy in the central region.

Our result is consistent with \citet{coelho11} in the sense that we
also find both young populations (and ongoing star formation) and old
populations in the bulge-region of our galaxies. This can be seen from
the lower-left panel of Figure~\ref{fig:slope_vs_center}: the barred
galaxies with no central turnover are located in the large-age and
flat-slope region, and the barred galaxies with turnover feature are
located in the region of younger ages and steeper slopes, forming an
anti-correlation of age versus slope. When plotted using the inward
extrapolation of the linear fit to the observed profile at larger
radii, the turnover galaxies move significantly toward regions of
older ages and flatter slopes, thus close to (thought not exactly the
same as) the non-turnover barred galaxies. This implies that, the
bulge which has stopped star formation long ago and is dominated by
very old populations ($\sim10$ Gyr) becomes actively forming stars and
younger, due to the ongoing/recent star formation induced by
bar-driven gas inflow. A direct comparison with \citet{coelho11}  is
not applicable, and for this one needs to perform stellar population
synthesis to the IFU spectra in order to decompose the young
population contributed by the bar-driven star formation from the
underlying old population. This can well be a topic for future
IFU-based studies.

\section{CONCLUSIONS} 
\label{sec:conclusions}

For a sample of 57 nearly face-on spiral galaxies selected from the
CALIFA/DR2 sample, we have performed photometric decomposition of
their bulge/bar/disk components using optical image from SDSS, and
obtained two-dimensional maps and radial profiles of \dindex,
\ewhda\ and \ewha\ using the integral field spectroscopy  from
CALIFA. We identify a class of ``turnover'' galaxies whose inner-most
region shows significant drop in \dindex, and/or corresponding upturn
in \ewhda\ and \ewha. We investigate the recent star formation
history, as indicated by the three diagnostic parameters, for the
central region of the both turnover and non-turnover galaxies, and  we
compare the results for barred and unbarred populations, as well as
for galaxies with different global properties.

Our conclusions can be summarized as follows.

\begin{enumerate}

\item We find strong link between the central turnover feature with
  the bar structure in galaxies. Out of the 57 galaxies in our sample,
  17 are identified as a turnover galaxy, of which 15 are barred. On
  the other hand, however, only half of the barred galaxies present
  the central turnover, indicating that the presence of a bar is a
  necessary, but not sufficient condition for the turnover feature to
  occur, at least for the sample being studied. 

\item The majority of the turnover galaxies identified by the central
  turnover in \dindex\ also present corresponding turnover feature  in
  the profiles of \ewha\ and \ewhda. Both the observed values of these
  diagnostic parameters and the values from inward extrapolation  of
  the linear fits to the profiles at larger radii are consistent with
  models in which star formation declines continuously, suggesting
  that  the  central region of these galaxies have been forming stars
  continuously in the past 1-2 Gyr. 

\item The turnover galaxies are found to have intermediate
  \nuvr\ colors ($3<$\nuvr$<5$), while their centers are mostly
  classified as star-forming regions according to both the central
  \dindex\ and the BPT diagram. When using the measurements from
  inward extrapolation of radial profiles at larger radii, these
  galaxies move largely towards the quiescent sequence with
  \dindex$>1.6$ and the AGN or AGN/SF composite regions.

\item In addition to the presence of the bar structure, the size of
  the bar (normalized by the disk radius) is the only galaxy property
  that is found to correlate with the turnover feature: the longer the
  bar, the larger the turnover region. There is no correlation of
  turnover size with galaxy stellar mass, \nuvr\ color, or the axial
  ratio of the bar.

\item Our results provide strong support to the expectations that
  bar-driven gas inflow triggers star formation in galaxy centers,
  which makes/grows/rejuvenates pseudobulges. 

\end{enumerate}

Our work demonstrates the power of integral field spectroscopy for
studying the resolved star formation history of galaxies. The
complementary photometric decomposition is also helpful and even
necessary in this work. A lesson we have learned is that, one may
easily mis-identify the turnover feature due to incorrect radial range
adopted for the linear fitting. The bulge/bar radii obtained from the
photometric decomposition have allowed us to take data well within the
inner region, avoiding contamination from the spiral disk.

We would like to point out that, however, the current work is limited
by the small sample size of the CALIFA survey. The exclusion of
merging systems from our sample is another thing one should keep in
mind.  A more complete sample covering wide ranges of galaxy mass,
morphology and environment is needed in order for a full understanding
of the central turnover. Such data is becoming available thanks to the
SDSS-IV/MaNGA survey \citep{bundy15}. We plan to extend  our analysis
to a much larger sample based on the MaNGA data, including galaxies
with masses down to $10^{9}$M$_\odot$ and those in merging/interacting
systems, and examining the connections with both the internal
structural properties such as the bar and external environment. 

\acknowledgments

We thank the anonymous referee for his or her careful and helpful comments,
which improved the quality of the paper immensely.
This work is supported by National Key Basic Research Program of China
(No. 2015CB857004), NSFC (Grant No. 11173045, 11233005, 11325314,
11320101002) and the Strategic Priority Research Program ``The
Emergence of Cosmological Structures'' of CAS (Grant No. XDB09000000).
Data used in this study comes from the Calar Alto Legacy Integral
Field Area (CALIFA) survey (http://califa.caha.es/) and the Sloan
Digital Sky Survey (SDSS). Based on observations collected at the
Centro Astron\'omico Hispano Alem\'an (CAHA) at Calar Alto, operated
jointly by the Max-Planck-Institut f\"ur Astronomie and the Instituto
de Astrof\'isica de Andaluc\'ia (CSIC).

Funding for  the SDSS and SDSS-II  has been provided by  the Alfred P.
Sloan Foundation, the Participating Institutions, the National Science
Foundation, the  U.S.  Department of Energy,  the National Aeronautics
and Space Administration, the  Japanese Monbukagakusho, the Max Planck
Society,  and the Higher  Education Funding  Council for  England. The
SDSS Web  Site is  http://www.sdss.org/.  The SDSS  is managed  by the
Astrophysical    Research    Consortium    for    the    Participating
Institutions. The  Participating Institutions are  the American Museum
of  Natural History,  Astrophysical Institute  Potsdam,  University of
Basel,  University  of  Cambridge,  Case Western  Reserve  University,
University of Chicago, Drexel  University, Fermilab, the Institute for
Advanced   Study,  the  Japan   Participation  Group,   Johns  Hopkins
University, the  Joint Institute  for Nuclear Astrophysics,  the Kavli
Institute  for   Particle  Astrophysics  and   Cosmology,  the  Korean
Scientist Group, the Chinese  Academy of Sciences (LAMOST), Los Alamos
National  Laboratory, the  Max-Planck-Institute for  Astronomy (MPIA),
the  Max-Planck-Institute  for Astrophysics  (MPA),  New Mexico  State
University,   Ohio  State   University,   University  of   Pittsburgh,
University  of  Portsmouth, Princeton  University,  the United  States
Naval Observatory, and the University of Washington.


\clearpage

\input{table1.tex}
\input{table2.tex}

\input{table3.tex}

\clearpage
\global\pdfpageattr\expandafter{\the\pdfpageattr/Rotate 90}

\end{document}

%% file: table1.tex
\begin{deluxetable}{lcccrrrrrcccc}
\tabletypesize{\tiny}
\tablecaption{The photometric parameters of bulge, bar and disk.}
\tablewidth{0pt}
\tablehead{
\colhead{Galaxy} & \colhead{Hubble} &
\colhead{Bar} & \colhead{Bar} &
\colhead{$\mu_{e}$} & \colhead{$r_{e}$} & \colhead{$n$} & 
\colhead{$\mu_{0}$} & \colhead{$h$} & 
\colhead{$\mu_{e,bar}$} & \colhead{$r_{e,bar}$} & \colhead{$n_{bar}$} & \colhead{$\chi^{2}$}\\
\colhead{} & \colhead{Type} &
\colhead{Type\tablenotemark{a}} &
\colhead{Type\tablenotemark{b}} &
\colhead{(mag arcsec$^{-2}$)} & \colhead{(arcsec)} & \colhead{} &
\colhead{(mag arcsec$^{-2}$)} & \colhead{(arcsec)} &
\colhead{(mag arcsec$^{-2}$)} & \colhead{(arcsec)} & \colhead{} & \colhead{}\\
\colhead{1} & \colhead{2} & \colhead{3} & \colhead{4} &
\colhead{5} & \colhead{6} & \colhead{7} & \colhead{8} & \colhead{9} & 
\colhead{10} & \colhead{11} & \colhead{12}  & \colhead{13}
}
\startdata
  UGC00005 & Sbc &  A & A & 22.80$\pm$0.09 & 1.26$\pm$0.01 & 0.61$\pm$0.28 & 21.74$\pm$0.18 &  7.83$\pm$1.00 &     \nodata    &    \nodata    &    \nodata    & 1.90 \\
   NGC7819 &  Sc &  A & A & 21.78$\pm$0.02 & 2.04$\pm$0.04 & 0.79$\pm$0.09 & 22.95$\pm$0.01 & 12.43$\pm$0.03 &     \nodata    &    \nodata    &    \nodata    & 1.37 \\
  UGC00036 & Sab & AB & A & 20.85$\pm$0.00 & 1.16$\pm$0.03 & 1.05$\pm$0.15 & 21.04$\pm$0.02 &  5.36$\pm$0.09 &     \nodata    &    \nodata    &    \nodata    & 1.23 \\
   NGC0001 & Sbc &  A & A & 20.04$\pm$0.00 & 1.30$\pm$0.01 & 0.68$\pm$0.16 & 20.87$\pm$0.01 &  5.89$\pm$0.02 &     \nodata    &    \nodata    &    \nodata    & 1.31 \\
   NGC0036 &  Sb &  B & B & 20.50$\pm$0.01 & 0.71$\pm$0.02 & 0.63$\pm$0.01 & 21.78$\pm$0.01 &  9.87$\pm$0.07 & 21.94$\pm$0.01 & 2.40$\pm$0.01 & 0.60$\pm$0.03 & 1.21 \\
   NGC0171 &  Sb &  B & B & 21.16$\pm$0.01 & 1.55$\pm$0.01 & 0.68$\pm$0.01 & 21.84$\pm$0.01 & 13.96$\pm$0.05 & 22.31$\pm$0.01 & 4.08$\pm$0.00 & 0.43$\pm$0.01 & 2.23 \\
   NGC0180 &  Sb &  B & B & 21.61$\pm$0.00 & 1.30$\pm$0.00 & 0.66$\pm$0.02 & 22.50$\pm$0.01 & 17.48$\pm$0.08 & 22.48$\pm$0.00 & 3.34$\pm$0.00 & 0.82$\pm$0.12 & 0.82 \\
   NGC0237 &  Sc &  B & A & 21.30$\pm$0.01 & 0.32$\pm$0.01 & 0.17$\pm$0.05 & 20.78$\pm$0.00 &  6.25$\pm$0.03 &     \nodata    &    \nodata    &    \nodata    & 1.13 \\
   NGC0477 & Sbc & AB & B & 22.45$\pm$0.01 & 1.19$\pm$0.00 & 0.88$\pm$0.10 & 22.50$\pm$0.01 & 11.65$\pm$0.86 & 22.79$\pm$0.01 & 2.15$\pm$0.01 & 0.38$\pm$0.08 & 1.45 \\
    IC1683 &  Sb & AB & A & 20.40$\pm$0.01 & 1.10$\pm$0.03 & 0.62$\pm$0.27 & 21.56$\pm$0.01 &  6.61$\pm$0.13 &     \nodata    &    \nodata    &    \nodata    & 1.07 \\
   NGC0496 & Scd &  A & A & 22.77$\pm$0.02 & 1.11$\pm$0.02 & 0.72$\pm$0.19 & 22.30$\pm$0.05 &  9.77$\pm$1.47 &     \nodata    &    \nodata    &    \nodata    & 0.95 \\
   NGC0776 &  Sb &  B & B & 21.66$\pm$0.00 & 1.66$\pm$0.02 & 0.41$\pm$0.01 & 21.89$\pm$0.01 & 10.63$\pm$0.12 & 21.40$\pm$0.06 & 1.92$\pm$0.07 & 1.10$\pm$0.28 & 1.52 \\
  UGC03253 &  Sb &  B & B & 21.20$\pm$0.00 & 0.93$\pm$0.00 & 0.62$\pm$0.02 & 21.81$\pm$0.01 &  8.47$\pm$0.38 & 22.35$\pm$0.02 & 2.10$\pm$0.00 & 0.43$\pm$0.09 & 1.29 \\
   NGC2253 & Sbc &  B & B & 21.43$\pm$0.02 & 0.97$\pm$0.02 & 0.17$\pm$0.03 & 21.07$\pm$0.01 &  8.57$\pm$0.07 & 20.54$\pm$0.01 & 0.96$\pm$0.00 & 1.27$\pm$0.01 & 1.92 \\
   NGC2347 & Sbc & AB & A & 20.77$\pm$0.00 & 1.26$\pm$0.00 & 0.86$\pm$0.00 & 20.52$\pm$0.00 &  6.65$\pm$0.00 &     \nodata    &    \nodata    &    \nodata    & 1.21 \\
   NGC2449 & Sab & AB & B & 23.06$\pm$0.01 & 1.12$\pm$0.01 & 0.17$\pm$0.12 & 21.12$\pm$0.03 &  7.04$\pm$0.06 & 21.22$\pm$0.00 & 1.57$\pm$0.00 & 0.71$\pm$0.11 & 1.41 \\
   NGC2730 & Scd &  B & B & 23.31$\pm$0.01 & 1.07$\pm$0.01 & 0.43$\pm$0.07 & 22.81$\pm$0.01 & 14.28$\pm$0.26 & 23.65$\pm$0.00 & 2.83$\pm$0.00 & 0.71$\pm$0.10 & 1.20 \\
   NGC2906 & Sbc &  A & A & 20.61$\pm$0.00 & 1.41$\pm$0.00 & 0.85$\pm$0.01 & 20.96$\pm$0.00 &  9.17$\pm$0.03 &     \nodata    &    \nodata    &    \nodata    & 2.02 \\
   NGC2916 & Sbc &  A & A & 20.17$\pm$0.02 & 1.06$\pm$0.04 & 0.51$\pm$0.03 & 21.16$\pm$0.01 &  8.77$\pm$0.02 &     \nodata    &    \nodata    &    \nodata    & 1.94 \\
  UGC05108 &  Sb &  B & B & 21.09$\pm$0.01 & 0.81$\pm$0.01 & 0.63$\pm$0.02 & 21.26$\pm$0.01 &  4.99$\pm$0.03 & 22.43$\pm$0.01 & 1.39$\pm$0.01 & 0.17$\pm$0.21 & 1.36 \\
   NGC3106 & Sab &  A & A & 20.90$\pm$0.01 & 2.51$\pm$0.02 & 1.72$\pm$0.03 & 22.36$\pm$0.12 & 14.39$\pm$0.31 &     \nodata    &    \nodata    &    \nodata    & 1.27 \\
   NGC3381 &  Sd &  B & B & 21.82$\pm$0.01 & 1.04$\pm$0.02 & 1.05$\pm$0.12 & 22.22$\pm$0.02 & 14.25$\pm$0.36 & 22.59$\pm$0.01 & 2.88$\pm$0.01 & 0.34$\pm$0.04 & 1.24 \\
    IC0674 & Sab &  B & B & 20.62$\pm$0.02 & 1.01$\pm$0.01 & 0.65$\pm$0.01 & 21.43$\pm$0.01 &  6.01$\pm$0.04 & 22.24$\pm$0.01 & 2.26$\pm$0.02 & 0.28$\pm$0.01 & 1.21 \\
   NGC3614 & Sbc & AB & A & 23.17$\pm$0.01 & 3.35$\pm$0.01 & 1.12$\pm$0.12 & 22.60$\pm$0.02 & 21.68$\pm$0.10 &     \nodata    &    \nodata    &    \nodata    & 0.62 \\
   NGC3811 & Sbc &  B & B & 25.96$\pm$0.02 & 3.41$\pm$0.04 & 1.42$\pm$0.01 & 21.84$\pm$0.04 & 12.84$\pm$0.23 & 20.53$\pm$0.02 & 1.80$\pm$0.10 & 1.11$\pm$0.05 & 1.95 \\
  UGC07012 & Scd & AB & A & 23.27$\pm$0.00 & 3.91$\pm$0.01 & 1.02$\pm$0.03 & 22.50$\pm$0.03 &  8.55$\pm$0.16 &     \nodata    &    \nodata    &    \nodata    & 1.43 \\
   NGC4185 & Sbc & AB & A & 22.27$\pm$0.01 & 2.14$\pm$0.01 & 0.83$\pm$0.12 & 22.22$\pm$0.02 & 16.25$\pm$0.10 &     \nodata    &    \nodata    &    \nodata    & 1.04 \\
   NGC4210 &  Sb &  B & B & 21.82$\pm$0.02 & 0.93$\pm$0.01 & 0.60$\pm$0.11 & 21.77$\pm$0.01 & 12.90$\pm$0.05 & 21.87$\pm$0.01 & 2.04$\pm$0.00 & 0.46$\pm$0.02 & 2.81 \\
   NGC4470 &  Sc &  A & A & 23.86$\pm$0.00 & 1.07$\pm$0.01 & 0.84$\pm$0.02 & 21.25$\pm$0.01 & 20.08$\pm$0.06 &     \nodata    &    \nodata    &    \nodata    & 1.58 \\
   NGC5000 & Sbc &  B & B & 21.31$\pm$0.01 & 0.80$\pm$0.01 & 0.51$\pm$0.05 & 22.61$\pm$0.00 & 11.37$\pm$0.05 & 23.42$\pm$0.00 & 4.46$\pm$0.00 & 1.13$\pm$0.14 & 1.20 \\
   NGC5205 & Sbc &  B & A & 21.60$\pm$0.00 & 2.17$\pm$0.01 & 0.83$\pm$0.53 & 21.77$\pm$0.01 & 10.77$\pm$0.08 &     \nodata    &    \nodata    &    \nodata    & 1.38 \\
  UGC08781 &  Sb &  B & B & 20.48$\pm$0.02 & 0.96$\pm$0.01 & 0.99$\pm$0.07 & 21.98$\pm$0.01 &  8.80$\pm$0.05 & 22.03$\pm$0.01 & 2.43$\pm$0.00 & 0.67$\pm$0.01 & 1.57 \\
   NGC5378 &  Sb &  B & B & 20.51$\pm$0.07 & 1.30$\pm$0.14 & 1.01$\pm$0.08 & 22.47$\pm$0.00 & 15.83$\pm$0.06 & 22.65$\pm$0.04 & 4.42$\pm$0.06 & 0.52$\pm$0.02 & 1.05 \\
   NGC5406 &  Sb &  B & B & 20.29$\pm$0.00 & 1.13$\pm$0.00 & 0.51$\pm$0.01 & 21.54$\pm$0.01 & 13.16$\pm$0.05 & 21.85$\pm$0.01 & 2.92$\pm$0.00 & 1.19$\pm$0.06 & 1.29 \\
   NGC5520 & Sbc &  A & A & 20.29$\pm$0.01 & 1.44$\pm$0.02 & 0.53$\pm$0.04 & 20.88$\pm$0.01 &  6.92$\pm$0.01 &     \nodata    &    \nodata    &    \nodata    & 1.85 \\
   NGC5720 & Sbc &  B & B & 20.95$\pm$0.02 & 1.16$\pm$0.02 & 0.77$\pm$0.12 & 22.36$\pm$0.01 & 10.09$\pm$0.06 & 22.52$\pm$0.01 & 2.50$\pm$0.01 & 0.45$\pm$0.01 & 1.11 \\
  UGC09476 & Sbc &  A & A & 22.45$\pm$0.01 & 1.54$\pm$0.01 & 0.69$\pm$0.12 & 22.17$\pm$0.02 & 11.23$\pm$0.10 &     \nodata    &    \nodata    &    \nodata    & 2.12 \\
   NGC5888 &  Sb &  B & B & 21.62$\pm$0.01 & 1.41$\pm$0.01 & 0.71$\pm$0.02 & 21.48$\pm$0.00 & 10.35$\pm$0.04 & 23.39$\pm$0.01 & 3.20$\pm$0.01 & 0.92$\pm$0.21 & 1.25 \\
    IC4566 &  Sb &  B & B & 21.24$\pm$0.02 & 1.86$\pm$0.01 & 0.91$\pm$0.11 & 22.11$\pm$0.01 & 12.08$\pm$0.04 & 23.49$\pm$0.01 & 2.20$\pm$0.01 & 0.17$\pm$0.02 & 1.34 \\
   NGC6004 & Sbc &  B & B & 20.72$\pm$0.09 & 0.93$\pm$0.19 & 0.64$\pm$0.10 & 22.12$\pm$0.01 & 14.03$\pm$0.19 & 22.99$\pm$0.04 & 2.45$\pm$0.05 & 0.58$\pm$0.20 & 1.67 \\
   NGC6063 & Sbc &  A & A & 22.99$\pm$0.07 & 1.20$\pm$0.21 & 0.78$\pm$0.04 & 22.28$\pm$0.00 & 11.38$\pm$0.06 &     \nodata    &    \nodata    &    \nodata    & 1.30 \\
   NGC6154 & Sab &  B & B & 26.99$\pm$0.01 & 3.95$\pm$0.02 & 0.17$\pm$0.11 & 22.23$\pm$0.01 & 11.76$\pm$0.03 & 21.49$\pm$0.01 & 2.24$\pm$0.02 & 1.43$\pm$0.01 & 1.23 \\
    IC1256 &  Sb & AB & B & 21.86$\pm$0.01 & 0.91$\pm$0.01 & 0.80$\pm$0.02 & 22.16$\pm$0.02 &  9.12$\pm$0.03 & 27.08$\pm$0.00 & 5.39$\pm$0.01 & 0.17$\pm$0.02 & 1.21 \\
   NGC6497 & Sab &  B & B & 20.36$\pm$0.01 & 1.07$\pm$0.01 & 1.03$\pm$0.02 & 21.51$\pm$0.01 &  7.67$\pm$0.04 & 22.40$\pm$0.00 & 2.62$\pm$0.03 & 0.32$\pm$0.08 & 1.26 \\
   NGC6941 &  Sb &  B & B & 20.81$\pm$0.01 & 1.39$\pm$0.03 & 0.78$\pm$0.04 & 22.49$\pm$0.01 & 19.48$\pm$0.03 & 22.50$\pm$0.01 & 4.38$\pm$0.00 & 0.41$\pm$0.14 & 1.32 \\
  UGC11649 & Sab &  B & B & 21.97$\pm$0.02 & 1.71$\pm$0.03 & 1.02$\pm$0.07 & 22.54$\pm$0.01 & 12.37$\pm$0.05 & 22.83$\pm$0.01 & 3.78$\pm$0.00 & 0.41$\pm$0.19 & 1.12 \\
   NGC7321 & Sbc &  B & B & 21.33$\pm$0.01 & 0.83$\pm$0.02 & 0.47$\pm$0.07 & 21.50$\pm$0.01 & 11.14$\pm$0.12 & 22.21$\pm$0.01 & 3.03$\pm$0.02 & 1.02$\pm$0.45 & 1.23 \\
  UGC12224 &  Sc &  A & A & 23.29$\pm$0.00 & 1.80$\pm$0.00 & 1.05$\pm$0.18 & 23.03$\pm$0.01 & 16.19$\pm$0.04 &     \nodata    &    \nodata    &    \nodata    & 0.97 \\
   NGC7466 & Sbc &  A & A & 20.97$\pm$0.02 & 1.59$\pm$0.01 & 0.75$\pm$0.01 & 22.14$\pm$0.00 & 12.35$\pm$0.05 &     \nodata    &    \nodata    &    \nodata    & 1.73 \\
   NGC7489 & Sbc &  A & A & 22.43$\pm$0.01 & 1.08$\pm$0.01 & 0.71$\pm$0.03 & 22.24$\pm$0.00 &  9.94$\pm$0.06 &     \nodata    &    \nodata    &    \nodata    & 1.24 \\
   NGC7549 & Sbc &  B & B & 23.51$\pm$0.02 & 1.45$\pm$0.00 & 0.86$\pm$0.08 & 21.69$\pm$0.00 &  8.41$\pm$0.04 & 23.01$\pm$0.62 & 2.30$\pm$0.83 & 0.65$\pm$0.15 & 1.55 \\
   NGC7563 &  Sa &  B & B & 19.95$\pm$0.00 & 1.01$\pm$0.00 & 0.89$\pm$0.05 & 21.92$\pm$0.01 & 10.71$\pm$0.02 & 21.14$\pm$0.00 & 3.66$\pm$0.00 & 0.74$\pm$0.13 & 1.22 \\
   NGC7591 & Sbc &  B & B & 20.49$\pm$0.01 & 0.33$\pm$0.01 & 0.36$\pm$0.08 & 20.72$\pm$0.01 &  6.02$\pm$0.02 & 21.41$\pm$0.11 & 0.34$\pm$0.36 & 0.17$\pm$0.06 & 2.24 \\
   NGC7653 &  Sb &  A & A & 20.87$\pm$0.01 & 1.95$\pm$0.01 & 1.17$\pm$0.32 & 21.61$\pm$0.07 &  9.57$\pm$0.08 &     \nodata    &    \nodata    &    \nodata    & 1.33 \\
   NGC7716 &  Sb &  A & A & 20.11$\pm$0.01 & 2.17$\pm$0.01 & 1.15$\pm$0.12 & 21.72$\pm$0.02 & 13.02$\pm$0.10 &     \nodata    &    \nodata    &    \nodata    & 2.18 \\
  UGC12816 &  Sc &  A & A & 23.22$\pm$0.01 & 2.53$\pm$0.01 & 1.04$\pm$0.58 & 22.65$\pm$0.07 &  8.93$\pm$0.13 &     \nodata    &    \nodata    &    \nodata    & 1.07 \\
   NGC5947 & Sbc &  B & B & 21.03$\pm$0.02 & 0.73$\pm$0.01 & 0.42$\pm$0.11 & 22.04$\pm$0.01 &  9.35$\pm$0.04 & 21.53$\pm$0.01 & 1.94$\pm$0.01 & 1.04$\pm$0.01 & 1.55
\enddata
\label{tbl1}
\footnotetext[1]{Bar classification by CALIFA team \citep{walcher14}: `A' for
unbarred, `B' for barred, `AB' if unsure.}
\footnotetext[2]{Our bar classification used in this work: `A' for unbarred,
`B' for barred. }
\end{deluxetable}

%% file: table2.tex
\begin{deluxetable}{lrrrrrrr}
\tabletypesize{\scriptsize}
\tablewidth{0pt}
\tablecaption{The Central Observed and Extrapolated Values and the Gradients.}
\tablehead{
& \multicolumn{3}{c}{D$_{n}$(4000)} 
& \multicolumn{3}{c}{SSP-AGE} \\
\cline{2-4} \cline{5-7} \\
\colhead{Galaxy} & \colhead{Cen$_{obs}$} & \colhead{Cen$_{extra}$} & \colhead{Slope(dex/R$_{e}$)} & 
\colhead{Cen$_{obs}$} & \colhead{Cen$_{extra}$} & \colhead{Slope(dex/R$_{e}$)} & \colhead{R$_{t}$($\arcsec$)}
}
\startdata
  UGC00005 &  1.58 $\pm$  0.09 &  1.57 $\pm$  0.10 & -0.40 $\pm$  0.30 &  9.43 $\pm$  0.15 &  9.50 $\pm$  0.28 & -1.01 $\pm$  0.87 & \nodata \\
   NGC7819 &  1.18 $\pm$  0.01 &  1.12 $\pm$  0.11 &  0.64 $\pm$  0.42 &  8.24 $\pm$  0.05 &  8.31 $\pm$  0.33 &  1.33 $\pm$  1.21 & \nodata \\
  UGC00036 &  1.82 $\pm$  0.12 &  1.85 $\pm$  0.22 & -0.15 $\pm$  0.42 &  9.65 $\pm$  0.05 &  9.70 $\pm$  0.21 & -0.06 $\pm$  0.37 & \nodata \\
   NGC0001 &  1.56 $\pm$  0.02 &  1.58 $\pm$  0.10 & -0.20 $\pm$  0.15 &  9.34 $\pm$  0.02 &  9.38 $\pm$  0.25 & -0.42 $\pm$  0.41 & \nodata \\
   NGC0036 &  1.86 $\pm$  0.05 &  2.09 $\pm$  0.28 & -0.74 $\pm$  0.72 &  9.78 $\pm$  0.02 & 10.05 $\pm$  0.41 & -1.02 $\pm$  1.09 &    3.00 \\
   NGC0171 &  1.85 $\pm$  0.19 &  1.93 $\pm$  0.29 & -0.38 $\pm$  0.87 &  9.85 $\pm$  0.25 & 10.04 $\pm$  0.36 & -0.73 $\pm$  1.03 & \nodata \\
   NGC0180 &  1.33 $\pm$  0.08 &  1.81 $\pm$  0.62 & -0.16 $\pm$  1.79 &  8.97 $\pm$  0.23 &  9.82 $\pm$  0.49 & -0.55 $\pm$  1.38 &    5.00 \\
   NGC0237 &  1.47 $\pm$  0.03 &  1.44 $\pm$  0.18 & -0.21 $\pm$  0.46 &  9.27 $\pm$  0.03 &  9.23 $\pm$  0.61 & -0.63 $\pm$  1.56 & \nodata \\
   NGC0477 &  1.40 $\pm$  0.03 &  1.63 $\pm$  0.26 & -0.27 $\pm$  0.68 &  9.14 $\pm$  0.05 &  9.52 $\pm$  0.42 & -0.56 $\pm$  1.12 &    4.00 \\
    IC1683 &  1.36 $\pm$  0.03 &  1.30 $\pm$  0.24 &  0.36 $\pm$  0.51 &  8.92 $\pm$  0.04 &  8.86 $\pm$  0.59 &  0.61 $\pm$  1.21 & \nodata \\
   NGC0496 &  1.26 $\pm$  0.03 &  1.29 $\pm$  0.09 & -0.07 $\pm$  0.33 &  8.60 $\pm$  0.08 &  8.73 $\pm$  0.30 & -0.23 $\pm$  1.21 & \nodata \\
   NGC0776 &  1.49 $\pm$  0.03 &  1.82 $\pm$  0.42 & -0.52 $\pm$  0.98 &  9.24 $\pm$  0.03 &  9.72 $\pm$  0.73 & -0.72 $\pm$  1.78 &    5.00 \\
  UGC03253 &  1.47 $\pm$  0.06 &  1.73 $\pm$  0.17 & -0.18 $\pm$  0.38 &  9.25 $\pm$  0.12 &  9.73 $\pm$  0.27 & -0.46 $\pm$  0.61 &    3.00 \\
   NGC2253 &  1.37 $\pm$  0.02 &  1.61 $\pm$  0.08 & -0.31 $\pm$  0.25 &  9.04 $\pm$  0.05 &  9.58 $\pm$  0.20 & -0.83 $\pm$  0.64 &    3.00 \\
   NGC2347 &  1.70 $\pm$  0.04 &  1.68 $\pm$  0.19 & -0.65 $\pm$  0.47 &  9.57 $\pm$  0.06 &  9.62 $\pm$  0.40 & -1.38 $\pm$  1.05 & \nodata \\
   NGC2449 &  1.86 $\pm$  0.05 &  1.85 $\pm$  0.34 & -0.28 $\pm$  0.72 &  9.77 $\pm$  0.03 &  9.80 $\pm$  0.57 & -0.42 $\pm$  1.21 & \nodata \\
   NGC2730 &  1.36 $\pm$  0.02 &  1.33 $\pm$  0.13 & -0.12 $\pm$  0.49 &  8.85 $\pm$  0.10 &  8.86 $\pm$  0.62 & -0.52 $\pm$  2.37 & \nodata \\
   NGC2906 &  1.89 $\pm$  0.00 &  1.96 $\pm$  0.09 & -0.73 $\pm$  0.37 &  9.76 $\pm$  0.01 &  9.83 $\pm$  0.16 & -0.59 $\pm$  0.57 & \nodata \\
   NGC2916 &  1.75 $\pm$  0.08 &  1.93 $\pm$  0.28 & -0.54 $\pm$  0.76 &  9.84 $\pm$  0.09 &  9.86 $\pm$  0.26 & -0.61 $\pm$  0.75 &    3.00 \\
  UGC05108 &  1.67 $\pm$  0.07 &  1.85 $\pm$  0.22 & -0.27 $\pm$  0.34 &  9.59 $\pm$  0.08 &  9.85 $\pm$  0.26 & -0.46 $\pm$  0.45 &    3.00 \\
   NGC3106 &  1.95 $\pm$  0.03 &  1.99 $\pm$  0.24 & -0.49 $\pm$  0.85 &  9.91 $\pm$  0.02 & 10.05 $\pm$  0.43 & -0.63 $\pm$  1.44 & \nodata \\
   NGC3381 &  1.15 $\pm$  0.05 &  1.30 $\pm$  0.09 & -0.11 $\pm$  0.31 &  8.06 $\pm$  0.29 &  8.74 $\pm$  0.39 & -0.39 $\pm$  1.36 &    3.00 \\
    IC0674 &  2.00 $\pm$  0.01 &  1.95 $\pm$  0.15 & -0.22 $\pm$  0.23 &  9.96 $\pm$  0.02 & 10.12 $\pm$  0.15 & -0.51 $\pm$  0.26 & \nodata \\
   NGC3614 &  1.55 $\pm$  0.02 &  1.76 $\pm$  0.27 & -0.87 $\pm$  1.04 &  9.42 $\pm$  0.06 &  9.73 $\pm$  0.51 & -1.56 $\pm$  2.11 &    4.00 \\
   NGC3811 &  1.46 $\pm$  0.07 &  1.78 $\pm$  0.27 & -0.45 $\pm$  0.59 &  9.14 $\pm$  0.07 &  9.93 $\pm$  0.38 & -0.99 $\pm$  0.90 &    5.00 \\
  UGC07012 &  1.29 $\pm$  0.05 &  1.31 $\pm$  0.09 & -0.11 $\pm$  0.23 &  8.63 $\pm$  0.19 &  8.77 $\pm$  0.35 & -0.42 $\pm$  0.94 & \nodata \\
   NGC4185 &  1.74 $\pm$  0.00 &  1.77 $\pm$  0.08 & -0.66 $\pm$  0.41 &  9.80 $\pm$  0.03 &  9.86 $\pm$  0.20 & -1.53 $\pm$  0.96 & \nodata \\
   NGC4210 &  1.71 $\pm$  0.09 &  1.91 $\pm$  0.12 & -0.75 $\pm$  0.45 &  9.74 $\pm$  0.14 &  9.95 $\pm$  0.23 & -1.17 $\pm$  0.81 &    3.00 \\
   NGC4470 &  1.28 $\pm$  0.02 &  1.30 $\pm$  0.07 & -0.05 $\pm$  0.21 &  8.71 $\pm$  0.07 &  8.84 $\pm$  0.26 & -0.41 $\pm$  0.79 & \nodata \\
   NGC5000 &  1.43 $\pm$  0.02 &  1.76 $\pm$  0.27 & -0.26 $\pm$  0.49 &  9.10 $\pm$  0.02 &  9.65 $\pm$  0.39 & -0.34 $\pm$  0.72 &    3.00 \\
   NGC5205 &  1.75 $\pm$  0.04 &  1.80 $\pm$  0.11 & -0.37 $\pm$  0.27 &  9.69 $\pm$  0.05 &  9.75 $\pm$  0.26 & -0.46 $\pm$  0.69 & \nodata \\
  UGC08781 &  1.90 $\pm$  0.18 &  1.96 $\pm$  0.11 & -0.22 $\pm$  0.20 &  9.84 $\pm$  0.09 & 10.05 $\pm$  0.18 & -0.42 $\pm$  0.33 & \nodata \\
   NGC5378 &  1.90 $\pm$  0.03 &  1.98 $\pm$  0.47 & -0.25 $\pm$  0.97 &  9.96 $\pm$  0.02 &  9.87 $\pm$  0.72 & -0.03 $\pm$  1.44 & \nodata \\
   NGC5406 &  1.97 $\pm$  0.02 &  1.98 $\pm$  0.19 & -0.34 $\pm$  0.61 &  9.87 $\pm$  0.04 &  9.89 $\pm$  0.17 & -0.48 $\pm$  0.53 & \nodata \\
   NGC5520 &  1.39 $\pm$  0.11 &  1.39 $\pm$  0.21 & -0.22 $\pm$  0.47 &  9.09 $\pm$  0.29 &  9.02 $\pm$  0.85 & -0.58 $\pm$  1.96 & \nodata \\
   NGC5720 &  1.96 $\pm$  0.07 &  2.05 $\pm$  0.14 & -0.52 $\pm$  0.36 &  9.95 $\pm$  0.02 & 10.04 $\pm$  0.13 & -0.58 $\pm$  0.38 & \nodata \\
  UGC09476 &  1.43 $\pm$  0.03 &  1.49 $\pm$  0.10 & -0.44 $\pm$  0.39 &  9.14 $\pm$  0.05 &  9.34 $\pm$  0.34 & -1.22 $\pm$  1.47 & \nodata \\
   NGC5888 &  2.02 $\pm$  0.03 &  1.96 $\pm$  0.14 & -0.15 $\pm$  0.31 &  9.96 $\pm$  0.01 &  9.98 $\pm$  0.15 & -0.23 $\pm$  0.32 & \nodata \\
    IC4566 &  1.92 $\pm$  0.07 &  1.88 $\pm$  0.17 & -0.28 $\pm$  0.47 &  9.92 $\pm$  0.07 &  9.98 $\pm$  0.29 & -0.68 $\pm$  0.72 & \nodata \\
   NGC6004 &  1.46 $\pm$  0.02 &  1.65 $\pm$  0.10 & -0.48 $\pm$  0.42 &  9.16 $\pm$  0.04 &  9.63 $\pm$  0.22 & -1.22 $\pm$  0.87 &    3.00 \\
   NGC6063 &  1.59 $\pm$  0.02 &  1.56 $\pm$  0.09 & -0.06 $\pm$  0.32 &  9.52 $\pm$  0.05 &  9.60 $\pm$  0.31 & -0.78 $\pm$  1.01 & \nodata \\
   NGC6154 &  2.04 $\pm$  0.07 &  1.98 $\pm$  0.23 & -0.03 $\pm$  0.47 &  9.93 $\pm$  0.02 & 10.08 $\pm$  0.26 & -0.35 $\pm$  0.53 & \nodata \\
    IC1256 &  1.51 $\pm$  0.02 &  1.61 $\pm$  0.08 & -0.24 $\pm$  0.15 &  9.28 $\pm$  0.03 &  9.55 $\pm$  0.20 & -0.62 $\pm$  0.36 & \nodata \\
   NGC6497 &  1.64 $\pm$  0.02 &  1.69 $\pm$  0.11 & -0.13 $\pm$  0.23 &  9.57 $\pm$  0.05 &  9.72 $\pm$  0.39 & -0.39 $\pm$  0.77 & \nodata \\
   NGC6941 &  1.82 $\pm$  0.09 &  1.99 $\pm$  0.28 & -0.35 $\pm$  0.84 &  9.79 $\pm$  0.11 &  9.97 $\pm$  0.23 & -0.44 $\pm$  0.70 &    3.00 \\
  UGC11649 &  1.86 $\pm$  0.04 &  1.89 $\pm$  0.30 & -0.19 $\pm$  0.64 &  9.88 $\pm$  0.04 & 10.07 $\pm$  0.34 & -0.66 $\pm$  0.72 & \nodata \\
   NGC7321 &  1.73 $\pm$  0.05 &  2.04 $\pm$  0.16 & -0.78 $\pm$  0.29 &  9.61 $\pm$  0.05 & 10.19 $\pm$  0.32 & -1.40 $\pm$  0.56 &    4.00 \\
  UGC12224 &  1.45 $\pm$  0.04 &  1.51 $\pm$  0.14 & -0.35 $\pm$  0.69 &  9.09 $\pm$  0.07 &  9.28 $\pm$  0.29 & -0.63 $\pm$  1.45 & \nodata \\
   NGC7466 &  1.55 $\pm$  0.01 &  1.48 $\pm$  0.23 & -0.02 $\pm$  0.49 &  9.20 $\pm$  0.09 &  9.31 $\pm$  0.42 & -0.20 $\pm$  0.91 & \nodata \\
   NGC7489 &  1.30 $\pm$  0.00 &  1.30 $\pm$  0.17 &  0.04 $\pm$  0.58 &  8.83 $\pm$  0.01 &  8.79 $\pm$  0.37 &  0.07 $\pm$  1.21 & \nodata \\
   NGC7549 &  1.42 $\pm$  0.05 &  1.42 $\pm$  0.22 & -0.12 $\pm$  0.44 &  9.17 $\pm$  0.10 &  9.11 $\pm$  0.62 & -0.34 $\pm$  1.18 & \nodata \\
   NGC7563 &  2.05 $\pm$  0.03 &  2.04 $\pm$  0.48 & -0.06 $\pm$  0.41 &  9.92 $\pm$  0.02 & 10.01 $\pm$  0.43 & -0.09 $\pm$  0.36 & \nodata \\
   NGC7591 &  1.56 $\pm$  0.02 &  1.48 $\pm$  0.28 & -0.10 $\pm$  0.67 &  9.38 $\pm$  0.06 &  9.12 $\pm$  0.59 & -0.06 $\pm$  1.30 & \nodata \\
   NGC7653 &  1.68 $\pm$  0.07 &  1.75 $\pm$  0.30 & -0.89 $\pm$  0.72 &  9.61 $\pm$  0.13 & 10.04 $\pm$  0.59 & -2.34 $\pm$  1.52 & \nodata \\
   NGC7716 &  1.97 $\pm$  0.02 &  1.98 $\pm$  0.27 & -0.76 $\pm$  0.79 &  9.87 $\pm$  0.02 & 10.00 $\pm$  0.26 & -0.92 $\pm$  0.78 & \nodata \\
  UGC12816 &  1.33 $\pm$  0.01 &  1.31 $\pm$  0.05 & -0.11 $\pm$  0.17 &  8.96 $\pm$  0.05 &  8.95 $\pm$  0.29 & -0.59 $\pm$  0.87 & \nodata \\
   NGC5947 &  1.59 $\pm$  0.02 &  1.76 $\pm$  0.06 & -0.51 $\pm$  0.17 &  9.41 $\pm$  0.02 &  9.75 $\pm$  0.16 & -1.00 $\pm$  0.43 &    3.00
\enddata
\label{tbl2}
\end{deluxetable}

%% file: table3.tex
\renewcommand{\thefootnote}{\alph{footnote}}

\begin{deluxetable}{ccccc}
\tablecaption{Statictis of turnovers from different Profiles.}
\tablewidth{0pt}
\tablehead{
& & \multicolumn{3}{c}{N$_{t}$} \\
\cline{3-5}
\colhead{ } & \colhead{Total} & \colhead{D$_{n}$(4000)} & 
\colhead{EW(H$\delta$)\tablenotemark{a}} & \colhead{EW(H$\alpha$)\tablenotemark{b}} 
}
\startdata 
All galaxies &      57 &           17 & 15(14) & 22(15) \\
Bar     &           32 &           15 & 13(12) & 16(13) \\
unBar   &           25 &            2 &  2(2)  &  6(2)  
\enddata
\tablecomments{ \label{tbl3}}
\footnotetext[1]{The number in brackets means the number of galaxies reveal turnover in both EW(H$\delta$) and D$_{n}$(4000) profiles.}
\footnotetext[2]{The number in brackets means the number of galaxies reveal turnover in both EW(H$\alpha$) and D$_{n}$(4000) profiles. }
\end{deluxetable}